\documentclass[10pt,twocolumn,superscriptaddress,prb,showpacs,floatfix,aps]{revtex4-2}
\usepackage[utf8]{inputenc}
\usepackage{amsmath}
\usepackage{amssymb}
\usepackage{graphicx}
\usepackage{svg}
\usepackage{xspace}
\usepackage{mathtools}
\usepackage{amsfonts}
\usepackage{bm}
\usepackage{bbm}
\usepackage{dsfont}
\usepackage{soul}
\usepackage{comment}
\usepackage{physics}
\usepackage[normalem]{ulem}
\usepackage{babel}
\usepackage{xcolor}

\usepackage[final]{hyperref} 
\hypersetup{
	colorlinks=true,       
	linkcolor=blue,        
	citecolor=blue,        
	filecolor=magenta,     
	urlcolor=blue         
}

\usepackage{nicematrix} 
\NiceMatrixOptions{cell-space-limits=1pt}

\begin{document}

\title{Active quantum matter from monitored pure-state dynamics}
\author{Jacob F.\ Steiner}
\affiliation{Department of Physics and Institute for Quantum Information and Matter, California Institute of Technology, Pasadena, California 91125, USA}

\author{Felix von Oppen}
\affiliation{\mbox{Dahlem Center for Complex Quantum Systems, Fachbereich Physik, and Halle-Berlin-Regensburg}\\ Cluster of Excellence CCE, Freie Universit\"at Berlin, 14195 Berlin, Germany}

\author{Reinhold Egger}
\affiliation{Institut f\"ur Theoretische Physik, Heinrich-Heine-Universit\"at, D-40225 D\"usseldorf, Germany}


\begin{abstract}
Quantum many-body systems coupled to out-of-equilibrium reservoirs can behave as active matter and exhibit signs of flocking. However, the resulting steady states are highly mixed and carry only weak quantum signatures. We show that signatures of active matter also arise in ensembles of pure states undergoing monitored quantum dynamics. We consider a spinful Luttinger liquid subject to measurement processes that shuffle spin-up particles to the left and spin-down particles to the right. For weak monitoring strengths and ferromagnetic spin interactions, we find power-law quantum correlations between spin density and charge current, which we identify as a hallmark of active quantum matter. The monitoring plays a dual role, generating the quantum active correlations for weak strengths while driving a Berezinskii-Kosterlitz-Thouless (BKT) phase transition to a short-range correlated state at larger strengths. 
\end{abstract}

\maketitle 

\textit{Introduction.}---Classical active matter has attracted enormous attention due to its rich phenomenology and broad range of applicability \cite{marchetti2013hydrodynamics,bechinger2016active,ramaswamy2017active,shankar2022topological}. A representative subclass of active matter are interacting assemblies of self-propelled particles which spontaneously align their direction of motion  and form a  flock \cite{vicsek1995novel,solon2013revisiting,chate2020dry,benvegnen2022flocking,vansaders2026measurement,vrugt2025exactlyactivematter}. The self-propulsion allows these systems to bypass the Mermin-Wagner theorem and exhibit spontaneous breaking of a continuous symmetry in two spatial dimensions \cite{vicsek1995novel,toner1995long}. Even one-dimensional (1D) active Ising chains, while not technically long-range ordered, exhibit a transition from a disordered to a phase-separated regime that is characterized by finite-lifetime clusters of particles moving together on a disordered background \cite{solon2013revisiting,solon2015phase,chate2020dry}. 

It is natural to ask whether and how these phenomena translate to quantum systems. Very recently, first forays into this largely uncharted territory have been undertaken, broadly following two distinct approaches. First, a number of works aimed at establishing minimal structures required to endow a quantum particle with the ability to propel itself. These considered quantum walks \cite{yamagishi_proposal_2024}, classical noise \cite{adachi2022activity,antonov2025engineering,antonov2025modeling,kundu2025run}, or heat-to-motion conversion \cite{penner2025heat}. Active motion has also been investigated for defects in spin textures \cite{delser2023skyrmion,hardt2025propelling}. Second, collective effects of interacting self-propelled quantum particles were studied based on nonhermitian Hamiltonian \cite{adachi2022activity,takasan2024activity} as well as Lindbladian \cite{khasseh2023active,hanai2025photoinduced} dynamics. The models studied in Refs.~\cite{adachi2022activity,takasan2024activity,khasseh2023active} are closely related to the active Ising model of Ref.~\cite{solon2013revisiting}, describing spin-$\tfrac{1}{2}$ particles subject to incoherent directed hopping and spin alignment mechanisms. In particular, \citet{khasseh2023active} predict signatures of flocking to survive in the presence of sufficiently weak quantum spin-flip processes. 

These first steps towards a theory of active quantum matter leave key issues open. If  directed motion and alignment are due to incoherent (and hence essentially classical) processes, one expects low-purity mixed states at long times, so that quantum signatures are observable only as transient correlations.  Here we pose, and  answer in the affirmative, the question: Can active quantum matter be based solely on the fundamental ingredients of quantum many-body dynamics, namely unitary time evolution and quantum measurements \cite{wiseman2009quantum,fazio2025many}? The active nature of the system then originates from  quantum measurements. Importantly, this makes quantumness an essential ingredient as (weak) classical measurements do not influence the system dynamics and hence cannot induce directionality or self-propulsion. Thus, quantum measurements provide a general construction principle for active matter in the deep quantum limit. In fact, for a given measurement trajectory, the system remains in a pure quantum state at all times given a pure initial state.
We show that quantum active behavior can be realized based on this principle. In addition to the rich and interesting physics revealed by our case study, this opens the door to exploring quantum mechanisms of activity in a wide range of systems of different  dimensionalities, geometries, or quantum statistics, to mention just a few. 

We study a 1D quantum active model of spin-$\frac12$ fermions with ferromagnetic Ising coupling, similar to Ref.~\cite{khasseh2023active}, but employ only monitored quantum dynamics, i.e., unitary dynamics combined with weak continuous measurements \cite{wiseman2009quantum}. We design the measurements to generate directed hopping and thus self-propulsion. Quantum measurements allow energy and information exchange between individual particles and the detectors, thereby equipping our model with the defining characteristic of quantum active matter \cite{vrugt2025exactlyactivematter}.
The resulting pure-state dynamics exhibit features of quantum active motion manifested in quasi-long-range (algebraically decaying) quantum correlations between spin density and charge current. Initially, these correlations grow with increasing monitoring strength or ferromagnetic coupling. In this sense, the system forms an incipient quantum flock. 

Importantly, the monitoring plays a dual role. Beyond a critical strength, it drives a transition to a phase in which the correlations between spin density and charge current become short-ranged. This BKT transition  \cite{Gogolin1998} between phases with distinct levels of quantum active correlations is directly related to the charge sharpening transition in U(1)-symmetric monitored quantum many-body systems \cite{barratt2022field,poboiko2023theory,poboiko2025measurement,guo2025field}.  We discuss our monitoring scheme---implemented via nonhermitian jump operators that drive self-propulsion---and its competition with the unitary dynamics within an intuitive lattice model. We use a corresponding bosonized continuum theory, a nonhermitian spinful Luttinger liquid with a sine-Gordon nonlinearity, to gain analytical access to the BKT transition.
This builds on recent work on spinless models with monitoring of hermitian operators \cite{buchhold2015nonequilibrium,buchhold2021effective,mueller2025monitored}. 

Lattice models of generic U(1)-symmetric monitored systems  \cite{barratt2022field,poboiko2023theory,poboiko2025measurement,guo2025field} exhibit a transition between volume- and area-law entangled phases in addition to the 
charge-sharpening transition \cite{skinner2019measurement,li2018quantum,gullans2020dynamical,choi2019quantum,cao2019entanglement,jian2020measurement,bao2020theory,alberton2021entanglement,Fisher2023,fava2023nonlinear,lumia2024measurement}. Calculations indicate that this entanglement transition occurs at even stronger monitoring within the charge-sharp phase \cite{barratt2022field,poboiko2023theory,poboiko2025measurement,guo2025field}. By analogy, we expect the BKT transition to separate two distinct volume-law entangled phases, while the entanglement transition might be viewed as the boundary between quantum and classical active phases.

\begin{figure}
    \centering
    \includegraphics[width=0.9\linewidth]{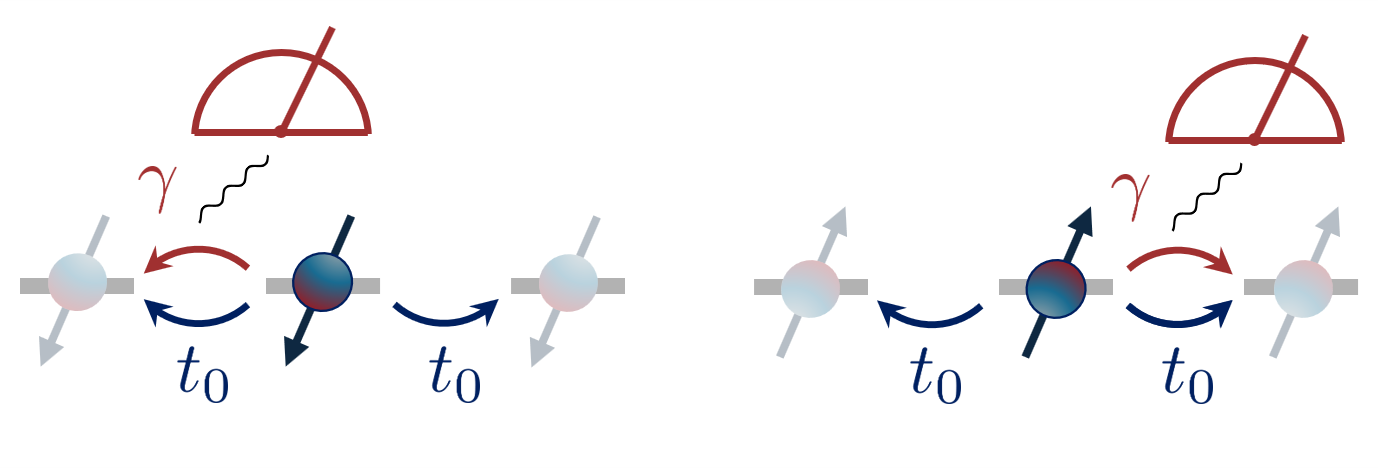}
    \caption{Cartoon of kinetic processes in the 1D monitored lattice model defined by Eqs.~(\ref{eq:sse})-(\ref{eq:lattice_jump_operators}).
    Spin-independent and bidirectional motion is due to hopping with amplitude $t_0$ (blue). Spin-dependent unidirectional motion is induced by monitoring with monitoring strength $\gamma$ (red). 
    }
    \label{fig:fig_1}
\end{figure}

\textit{Lattice model.}---We describe the monitored evolution of a 1D chain of spin-$\frac{1}{2}$ fermions $c_{x,\sigma}$ with site index $x\in\mathbb{Z}$ and spin
$\sigma\in\{\uparrow,\downarrow\}\equiv\{+1,-1\}$ by the stochastic Schrödinger equation 
\cite{wiseman2009quantum,fazio2025many,gross2018qubit}
\begin{multline}\label{eq:sse}
    i\frac{d}{dt}\ket{\psi_c} = \Big\{ H - \frac{i}{2} \sum_{x,\sigma} \Big[L^\dagger_{x,\sigma} L^{}_{x,\sigma} - 2W^{}_{x,\sigma} L^{}_{x,\sigma} + W^2_{x,\sigma}  \\
    + 2 \xi_{x,\sigma} (L^{}_{x,\sigma} - W^{}_{x,\sigma}) \Big]\Big\}  \ket{\psi_c}.
\end{multline}
The subscript $c$ denotes conditioning on the measurement trajectory, and the terms involving $W_{x,\sigma}^{} = \Re \bra{\psi_c} L_{x,\sigma}^{} \ket{\psi_c}$ ensure 
normalization of $\ket{\psi_c}$. The unitary dynamics obey a Hamiltonian $H$ including nearest-neighbor hopping $t_0$ and ferromagnetic Ising exchange $J_z>0$, 
\begin{equation}\label{eq:lattice_hamiltonian}
    H = \sum_{x}\bqty{- t_0 \pqty{c_x^\dagger  c^{}_{x+1} + {\rm h.c.}} -J_z S_x^z S_{x+1}^z  } + H_\textrm{flip},
\end{equation}
where $S_x^z=\frac12 c_x^\dagger \sigma_3 c_x$ and  $c_x = [c_{x,\uparrow},c_{x,\downarrow}]^\top$. The Pauli matrices ($\sigma_1,\sigma_2,\sigma_3$) act in spin space. 
We also allow for spin-nonconserving dynamics through a time-reversal odd Zeeman term, $H_{\textrm{flip}} = h \sum_x c_x^\dagger  \sigma_1 c^{}_{x}$,  or a time-reversal even Rashba spin-orbit coupling, $H_{\textrm{flip}} = -i \alpha \sum_{x} c_x^\dagger  \sigma_1 c^{}_{x+1} + {\rm h.c.}$. The self-propulsion of  spin-$\sigma$ fermions is achieved by monitoring with the nonhermitian jump operators
\begin{equation}\label{eq:lattice_jump_operators}
    L_{x,\sigma} = \sqrt{\gamma}\, c_{x+\sigma,\sigma}^\dagger c_{x,\sigma}^{},
\end{equation}
with monitoring strength $\gamma$. 
Physically, we continuously monitor jumps of a spin-$\sigma$ particle by one site in a spin-dependent direction, $x\to x+\sigma$. The random sequence of measurement outcomes is encoded by the Gaussian white noise sources $\xi_{x,\sigma}$ with 
$\overline{\xi_{x,\sigma}(t)}=0$ and correlations $\overline{\xi_{x,\sigma}(t)\xi_{x',\sigma'}(t')} = \delta_{xx'}\delta_{\sigma\sigma'}\delta(t-t')$,  where the overline denotes averaging over the ensemble of measurement outcomes. We emphasize that for a single measurement trajectory, Eq.~\eqref{eq:sse} preserves purity at all times $t$.

The kinetic processes of our model are graphically represented in Fig.\ \ref{fig:fig_1}. Spin-independent coherent hopping without preferred direction (amplitude $t_0$) builds up entanglement. Spin-dependent directional stochastic jumps are induced by monitoring (jump operators in Eq.\ \eqref{eq:lattice_jump_operators}). The monitoring detects jumps of spin-dependent directionality. In principle, this 
can be realized experimentally, e.g., by exploiting cotunneling by coupling the sites to an auxiliary chain of spin-polarized detector quantum dots \cite{Nazarov2009} and measuring their charge state, as explained in detail in the Supplemental Material (SM) \cite{SI}. Our model differs from the one discussed in Ref.~\cite{khasseh2023active} even with regard to the Lindblad evolution $\dot{\rho}_1  =  \mathcal{L}\rho_1= -i[H,\rho_1]+\sum_{x,\sigma} ( L\rho_1 L^\dagger - \tfrac{1}{2}\Bqty{L^\dagger L,\rho_1})$ of the ensemble-averaged state $\rho_{1} = \overline{\ketbra{\psi_c}{\psi_c}}$ \cite{wiseman2009quantum,fazio2025many,nielsen2010quantum}. First, we include spin-independent coherent hopping. Second, the Ising coupling enters the unitary dynamics. Third, we do not include an explicit alignment process. As a consequence, $\rho_1$ heats up to become completely mixed, $\rho_1(t \to \infty) \propto \mathbbm{1}$, suppressing signatures of quantum active motion in the ensemble-averaged steady state \footnote{While the $L_{x,\sigma}$ are not hermitian, one explicitly checks $\mathcal{L} \mathbbm{1} = \sum_{x} (n_{x+\sigma,\sigma} - n_{x,\sigma}) = 0$ with $n_{x,\sigma}=c^\dagger_{x,\sigma} c^{}_{x,\sigma}$ by translation invariance.}.
 
Importantly, signatures of quantum active motion persist at long times in the pure-state dynamics generated by Eq.~\eqref{eq:sse}. We consider measurement-averaged equal-time connected correlation functions,
\begin{equation}\label{eq:correlation_1}
    C_{AB}(x) =\overline{\bra{\psi_c} A_{x} B_0 \ket{\psi_c} - \bra{\psi_c}  A_{x} \ketbra{\psi_c} B_0  \ket{\psi_c} }.
\end{equation}
These exhibit nontrivial behavior due to the second term, which is nonlinear in the system's density matrix $\ketbra{\psi_c}$ \cite{buchhold2021effective}. Here, $A_x$ and $B_0$ represent local operators (charge and spin densities as well as currents) at sites $x$ and $0$, respectively. We express $C_{AB}(x)$ in terms of the two-replica density matrix averaged over measurement outcomes, $\rho_2 = \overline{\ketbra{\psi_c}{\psi_c} \otimes \ketbra{\psi_c}{\psi_c}}$. This yields
\begin{equation}\label{eq:correlation_2}
    C_{AB}(x) = \frac{1}{2} \ev{\pqty{A_{x}^{(1)}- A_{x}^{(2)}}\pqty{B_0^{(1)}- B_0^{(2)}} },
\end{equation}
where the superscript indicates action on replica $i=1,2$ and $\ev{\ldots} = \tr\pqty{\ldots \rho_2}$. 

\begin{figure*}[t]
    \centering
    \includegraphics[width=\linewidth]{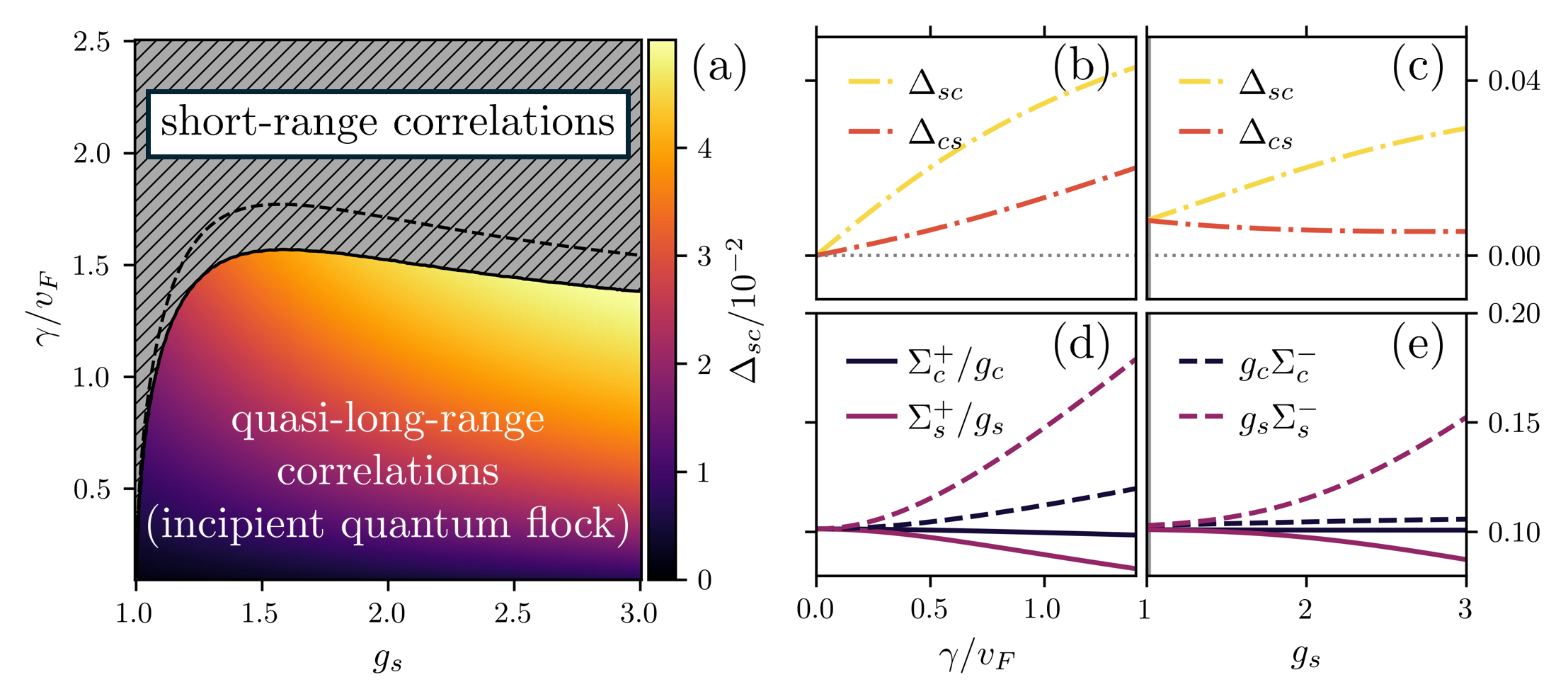}
    \caption{(a) Phase diagram in the $g_s$--$\gamma$ plane obtained from the RG equations \eqref{eq:RG_eqs} with $g_c=1$ and $\tilde c=\tilde s=1/\sqrt{2}$, see Eq.~\eqref{eq:couplings}. The quasi-long-range quantum active phase (small $\gamma$) and the short-range phase (large $\gamma$) are separated by a BKT transition (black line). In the quantum active phase, color  indicates the magnitude of $\Delta_{cs}$. 
    The dashed line is an analytical estimate for the phase boundary (see main text). (b-e) Dependence of the coefficients $\{\Sigma_{c/s}^\pm,\Delta_{cs/sc}\}$ on $\gamma$ and  $g_s$, which determine the correlations at the fixed point $\lambda=0$, see Eq.~\eqref{eq:correlationsfinal}, 
    using $g_s=2$ in (b) and (d), and $\gamma/v_F = 0.5$ in (c) and (e).}
    \label{fig:fig_2}
\end{figure*}

A natural signature for active self-propulsion is offered by the correlations between the local spin density (``magnetization'') $\varrho_s$ and the charge current density $j_c$.  In particular, such correlations are significant if spin-up (spin-down) particles move to the right (left) since a  spin-density imbalance is then 
accompanied by a charge current density.
We find that below a critical measurement rate, the corresponding steady-state correlations exhibit power-law behavior at long distances, $C_{\rho_s j_c}(x) \sim - \Delta_{sc}/x^2$. We emphasize that these are quantum correlations of a pure-state ensemble and are thus accompanied by significant entanglement. The coefficient $\Delta_{sc}$ is an increasing function of $\gamma$: activity is driven by the measurements. When $\gamma$ exceeds a critical strength, the pure-state dynamics undergoes a transition into a phase with exponentially decaying $C_{\rho_s j_c}(x)$. This is summarized in the phase diagram in Fig.~\ref{fig:fig_2}(a), where $g_s$ is the Luttinger liquid parameter quantifying the Ising interaction (see below). The color scale indicates 
the active correlations $\propto\Delta_{sc}$, which  increase with $\gamma$ right up to the transition, see Fig.~\ref{fig:fig_2}(b) for a linecut. 

\textit{Bosonization.}---Rather than working directly with the lattice model in Eqs.~\eqref{eq:lattice_hamiltonian} and \eqref{eq:lattice_jump_operators}, we consider a bosonized continuum theory, $H \to H_0 + \delta H$ and $L_{x,\sigma} \to L_\sigma (x)$, which captures this BKT transition and allows for analytical progress. (From now on, $x$ is a continuous coordinate.)
For generic incommensurate band fillings, the quadratic Hamiltonian
\begin{equation}\label{eq:bosonized_hamiltonian}
    H_0 = \frac{v_F}{2} \sum_\nu \int dx \bqty{(\partial_x \theta_\nu)^2 + \frac{1}{g^2_\nu} (\partial_x \varphi_\nu)^2}
\end{equation}
describes the dynamics of the charge ($\nu = c$) and spin ($\nu = s$) degrees of freedom in terms of two Luttinger liquids with Fermi velocity $v_F$. Here,  $\partial_x \theta_\nu(x)$ and $\varphi_\nu(x)$ are canonically conjugate real bosonic fields with $[\varphi_{\nu}(x),\partial_{x'} \theta_{\nu'}(x')] = i \delta_{\nu\nu'} \delta(x-x')$ \cite{Gogolin1998,egger1996,egger_scaling_1998}. 
Their physical meaning is explained after Eq.~\eqref{eq:bosonized_jump_operators_lin}, see also
Ref.~\cite{Gogolin1998}.
Measuring lengths in units of the lattice spacing, the spin sector has the Luttinger parameter $g_s \simeq 1/[1- J_z/2\pi v_F]^{1/2} > 1$, while $g_c = 1$ in the charge sector. 
Zeeman and Rashba terms as well as backscattering processes due to $J_z$ give rise to sine-Gordon terms collected in $\delta H$ (see SM \cite{SI}). Similarly, the bosonized measurement operators 
$L_\sigma(x) = M_\sigma(x) + i K_\sigma(x)$ with hermitian $M_\sigma$ and $K_\sigma$ decompose into linear and sine-Gordon terms, $M_{\sigma}=M_{0,\sigma}+\delta M_\sigma$ and $K_\sigma=K_{0,\sigma}+\delta K_\sigma$.  Here, \begin{subequations}\label{eq:bosonized_jump_operators_lin}
\begin{align}
    M_{0,\sigma} =&\ \sqrt{\frac{\gamma}{2\pi}} \tilde c\,  \partial_x \pqty{\varphi_c + \sigma \varphi_s} \sim \textrm{density}_\sigma,\\
    K_{0,\sigma} =&\ \sqrt{\frac{\gamma}{2\pi}} \tilde s\, \sigma\partial_x \pqty{\theta_c +  \sigma\theta_s} \sim \sigma\times  \textrm{current}_\sigma
\end{align}
\end{subequations}
stem from momentum-conserving processes. ($\tilde c$ and $\tilde s$ with $\tilde c^2 + \tilde s^2 = 1$ are real constants controlled by the band filling.) The hermitian part $M_{0,\sigma}$ depends on the smooth component $\partial_x \pqty{\varphi_c + \sigma \varphi_s}/\sqrt{2}$ of the density of spin-$\sigma$ particles. Similarly, the antihermitian part $iK_{0,\sigma}$ depends on the smooth component $\sigma \partial_x \pqty{\theta_c + \sigma \theta_s}/\sqrt{2}$ of the Hamiltonian current density in $\sigma$ direction due to spin-$\sigma$ particles. Further, $2k_F$ processes give rise to the sine-Gordon contributions
\begin{equation}\label{eq:bosonized_jump_operators_nonlin}
    \delta M_\sigma = \sqrt{\frac{\gamma}{\pi^2}} \sigma \sin\bqty{2k_F x + \sqrt{2\pi}\pqty{ \varphi_c + \sigma \varphi_s}} ,   
\end{equation}
while $\delta K_\sigma = 0$. Due to the large difference in momentum transfers, we can treat the $2k_F$ term as a separate measurement process \cite{buchhold2021effective}. Thus, there are two sets of jump operators, $L_{0,\sigma} =  M_{0,\sigma} + i K_{0,\sigma}$ and $\delta L_\sigma =  \delta M_\sigma$. In the bosonized theory, heating of $\rho_1$ manifests as unbounded growth of the correlation functions $\tr[\chi_{\pm,\nu}(x)\chi_{\pm',\nu'}(x)\rho_1(t)]$ for $t\to \infty$, where $\chi_{+,\nu} = \varphi_\nu$ and $\chi_{-,\nu} = \theta_\nu$ \cite{buchhold2015nonequilibrium}. 
We recall that $\rho_1$ is the ensemble-averaged state while $\rho_2$ refers to the averaged
two-replica density matrix.

\textit{Nonhermitian sine-Gordon theory.}---We now consider the time evolution of $\rho_2$. Correlations of charge and spin densities and/or currents can be traced to $C_{AB}(x)$ with $A,B \sim \partial_x \chi_{\pm,\nu}$, which  depends only on the relative modes $\chi^{(r)} = (\chi^{(1)} - \chi^{(2)})/\sqrt{2}$, see Eq.~\eqref{eq:correlation_2}. Conversely, as a consequence of the heating of $\rho_1$, correlations of the center-of-mass (absolute) modes $\chi^{(a)} = (\chi^{(1)} + \chi^{(2)})/\sqrt{2}$ grow without bound for $t\to \infty$. In fact, symmetry under replica exchange gives $\ev*{\chi^{(a)}} = \sqrt{2}  \tr(\chi\rho_1)$, $\ev*{\chi^{(r)}} = 0$, and
\begin{equation}
     \ev*{\chi^{(a)}_\alpha \chi^{(a)}_\beta }=\  2 \tr(\chi_\alpha \chi_\beta \rho_1) - \ev*{\chi^{(r)}_\alpha \chi^{(r)}_\beta}.
\end{equation}
Assuming that correlations of the relative fields remain finite (as confirmed a posteriori), unbounded heating of $\rho_1$ translates to unbounded heating of correlations in the absolute sector: $\ev*{\chi^{(a)}_\alpha(x) \chi^{(a)}_\beta(x)} \to \infty$ and $\ev*{\chi^{(a)}(x)} = 0$. To arrive at an effective description of the relative sector, we trace out the absolute modes, see SM \cite{SI} for details. Importantly, at long times, all  contributions from $\delta H$ become irrelevant under this procedure.  Using $\rho^{(r)}_2 = \tr^{(a)}\rho_2$, the dynamics in the relative replica sector is then governed by 
\begin{equation}\label{eq:eom_relative_sector}
    \dot{\rho}^{(r)}_2 = -i H_\textrm{eff}^{(r)} \rho^{(r)}_2 +i \rho^{(r)}_2 [H_\textrm{eff}^{(r)}]^\dagger.
\end{equation}
Mostly dropping the superscript $(r)$  from now on, the effective nonhermitian sine-Gordon Hamiltonian  $H_\textrm{eff} = v_F \int dx (\mathcal{H}_{\textrm{eff},0} +  \delta \mathcal{H}_{\textrm{eff} })$ is given by
\begin{subequations}\label{eq:eff_hamiltonian}
\begin{align}
    \mathcal{H}_{\textrm{eff},0} =&\   \frac{1}{2}  \sum_{\nu=c,s} \bqty{ \pqty{\partial_x \theta_\nu}^2 + (k_\nu+\kappa^2) \pqty{\partial_x \varphi_\nu}^2 }  \nonumber  \\ & +  \kappa \pqty{ \partial_x \varphi_c \partial_x \theta_s + \partial_x\varphi_s\partial_x\theta_c }, \\
    \delta  \mathcal{H}_{\textrm{eff} } =&\ i \lambda \cos\pqty{\sqrt{4\pi} \varphi_c} \cos\pqty{\sqrt{4\pi} \varphi_s},
\end{align}
\end{subequations}
where we introduced the couplings 
\begin{equation}\label{eq:couplings}   
   k_\nu = \frac{1}{g_\nu^{2}} - \frac{2i \tilde{c}^2\gamma}{\pi v_F}-\kappa^2,
   \quad \kappa = \frac{\tilde{c} \tilde{s} \gamma}{\pi v_F}, \quad \lambda = \frac{\gamma}{\pi^2 v_F}. 
\end{equation}  
Note that $\gamma$ enters both the parity-odd spin-charge coupling $\kappa$ (giving rise to $\Delta_{sc}\ne 0$) and the parity-even nonlinearity $\lambda$ (driving the BKT transition), highlighting the dual role of monitoring for quantum active matter.
Importantly, Eq.~\eqref{eq:eom_relative_sector} does not feature a jump term and thus describes the nonhermitian evolution of a pure state, $\rho_2^{(r)} = \ketbra*{\psi}$, towards a dark state $\ket*{\psi(t \to \infty)} = \ket*{0}$. Steady-state connected correlation functions are then given by $C_{AB}(x) = \frac{1}{2}\bra*{0} A^{(r)}(x) B^{(r)}(0) \ket*{0}$.  

\textit{Renormalization and phase diagram.}---To find $\ket*{0}$, we treat the nonlinear term $\propto \lambda$ in Eq.~\eqref{eq:eff_hamiltonian} using a perturbative renormalization group (RG) analysis \cite{Gogolin1998,buchhold2021effective}. The one-loop RG equations for the complex couplings \eqref{eq:couplings} are (for a derivation, see SM \cite{SI})
\begin{equation}\label{eq:RG_eqs}
    \frac{d{\lambda}}{d\ell} =  \pqty{2-\sum_\nu \frac{1}{\sqrt{k_\nu}}}{\lambda},\quad 
    \frac{d k_\nu}{d\ell}  = - f {\lambda}^2, 
\end{equation}
where $\ell$ is the standard RG flow parameter, $|f| \sim {\cal O}(1)$ depends weakly on $k_\nu$,
and $\kappa$ is not renormalized. Solving Eq.~\eqref{eq:RG_eqs} with the initial values \eqref{eq:couplings} gives the phase diagram  in Fig.~\ref{fig:fig_2}(a), with the solid curve indicating the transition between a phase with quasi-long-range quantum active correlations, where $\lambda(\ell\to \infty)=0$, 
and a phase with short-range correlations, where $\lambda$ flows towards strong coupling. The RG equations \eqref{eq:RG_eqs} are characteristic for a BKT  transition \cite{Gogolin1998,buchhold2021effective,mueller2025monitored}. Compared to the 2D classical case, the BKT transition occurs here in a $(1+1)$-dimensional quantum system.  In the quantum active phase, one may assume that $k_\nu$ remains unrenormalized since $\lambda$ quickly approaches zero. 
The resulting estimate for the phase boundary, $\sum_\nu 1/\sqrt{k_\nu}=2$, yields the dashed curve in Fig.~\ref{fig:fig_2}(a).  
We find that the quantum active phase disappears without   interactions ($g_s \to 1$). 

\textit{Gaussian fixed points.}---We now study the steady-state correlations $C_{AB}(x)$ for all combinations of $A,B \in \{\varrho_{c/s} ,j_{c/s}\}$ at the $\abs{\lambda} \to 0,\infty$ fixed points.  Since these limits correspond to Gaussian fixed points, we obtain exact results for the long-distance behavior. With the mean density $\varrho_0$, charge and spin densities are given by $\varrho_\nu=\sqrt{2/\pi}\partial_x\varphi_\nu+\varrho_0\delta_{\nu,c}$. The respective current densities $j_\nu$ include measurement-induced contributions $\propto \kappa$, see Refs.~\cite{SI,hovhannisyan2019quantum,popperl2023measurements,antonic2025motion},
\begin{equation}\label{eq:current_ops}
j_\nu=-\sqrt{{2}/{\pi}}v_F [\partial_x\theta_\nu - \kappa \partial_x\varphi_{\overline{\nu}}] + v_F\kappa\varrho_0\delta_{\nu,s},   
\end{equation}
with $\overline{\nu}=s,c$ for $\nu=c,s$.  Note that monitoring induces a constant spin-dependent  
current. Consider first the case $\lambda \to 0$, where we find  
\begin{equation} \label{eq:correlationsfinal}
    \begin{pmatrix}
        C_{\varrho^{}_\nu\varrho_{\nu'}}  & C_{\varrho^{}_\nu j_{\nu'}} \\
        C_{j^{}_\nu \varrho_{\nu'}}   & C_{j^{}_\nu j_{\nu'}}  
    \end{pmatrix} \simeq-  \frac{1}{x^2} 
    \begin{pmatrix}
        \delta_{\nu\nu'}  \Sigma^+_\nu & v_F \delta_{\nu\overline{\nu}'}\Delta_{\nu\overline{\nu}}\\
         v_F\delta_{\nu\overline{\nu}'}\Delta_{\overline{\nu}\nu} & v_F^2 \delta_{\nu\nu'} \Sigma^-_\nu
    \end{pmatrix}.
\end{equation}
The six real coefficients $\{\Sigma_{c/s}^\pm,\Delta_{cs/sc}\}$ are illustrated as functions of $\gamma$ and/or $g_s$ in Fig.~\ref{fig:fig_2}(b-e). As in equilibrium Luttinger liquids \cite{Gogolin1998}, density-density and current-current correlations decay $\propto x^{-2}$, implying that members of the pure-state ensemble are significantly entangled \cite{calabrese2004entanglement,buchhold2021effective}. Note that spin current (density) correlations are enhanced (reduced) by finite $\gamma$ and by $g_s > 1$, while charge correlations are not strongly affected. Equation~\eqref{eq:correlationsfinal} reveals quantum active correlations between spin (charge) densities and charge (spin) currents 
from the $\Delta_{sc (cs)}$ terms.
Notably,  $\Delta_{sc}$ is enhanced by the Ising interaction while $\Delta_{cs}$ is weakened,
which can be rationalized by noting that Ising interactions favor alignment of adjacent spins and thus quantum flocking.
Finally, consider the strong-coupling phase with large $|\lambda|$. Here, the nonlinearity in Eq.~\eqref{eq:eff_hamiltonian} leads to a Gaussian fixed point with a mass term in the boson dispersion (mass $m_\lambda$). 
We then find exponential decay, $C_{A_{\nu} B_{\nu'}}(x)\propto e^{-|x|/\xi_{\nu\nu'}}$, with the length scale \cite{SI}
\begin{equation}\label{eq:correlation_length}
    \xi_{\nu\nu'} \sim \frac{v_F}{\abs{m_\lambda}}\max\pqty{\sqrt{\abs{k_\nu}},\sqrt{\abs{k_{\nu'}}}}.
\end{equation} 
We thus encounter only short-range correlations. 

\textit{Discussion.}---We showed that quantum active matter can be realized in monitored interacting fermion chains. Monitoring causes self-propulsion but eventually also drives a BKT transition between phases with algebraic and short-ranged quantum active correlations. The algebraic phase is strongly quantum in that typical pure states cannot be prepared by shallow circuits. The short-range correlated phase arises from the quantum Zeno effect under frequent measurements. This phase is expected to host a second transition to a weakly quantum area-law entangled phase \cite{barratt2022field,poboiko2023theory,poboiko2025measurement,guo2025field}, with typical pure states preparable by shallow circuits \cite{suzuki2025quantum}. An explicit study of the entanglement  transition is left for future work, as it is currently unknown how to describe it within the bosonization approach \cite{mueller2025monitored}. Matrix product state methods may be useful in this context. Experimental studies of the quantum active correlations require many copies of the ensemble of measurement trajectories. To deal with the associated postselection problem \cite{Fisher2023}, one can use methods such as classical-quantum cross-correlations \cite{garratt2024probing}, classical postprocessing \cite{mcginley2024postselectionfree}, decoding \cite{dehghani2023neuralnetwork}, space-time duality \cite{ippoliti2021postselectionfree}, or tree-like measurements \cite{feng2026postselectionfree}. The entanglement transition was successfully observed in this way \cite{noel2022measurementinduced,Koh2023,Hoke2023}. We expect our work to stimulate theoretical and experimental research on active matter
in the deep quantum limit by employing quantum measurements.

\emph{Data availability.---}The raw data underlying the figures presented in this work are available at Zenodo \cite{Zenodo}.

\begin{acknowledgments}
    We thank Liliana Arrachea, Michael Buchhold, Sebastian Diehl, Rosario Fazio, Igor Gornyi, and Hartmut L\"owen for valuable discussions.  We acknowledge funding by the Deutsche Forschungsgemeinschaft (DFG, German Research Foundation) under Projektnummer 277101999 - TRR 183 (project B02), under Projektnummer EG 96/14-1, and under Germany's Excellence Strategy - Cluster of Excellence Matter and Light for Quantum Computing (ML4Q) EXC 2004/2 - 390534769.  J.F.S.~acknowledges the support of the AFOSR MURI program, under Agreement No. FA9550-22-1-0339.
\end{acknowledgments}

\bibliography{refs_draft}

@misc{Zenodo,
author={J.F. Steiner and F. von Oppen and R. Egger},
title={Active quantum matter from monitored pure-state dynamics  [Raw data]},
url={https://zenodo.org/records/21270406}
}

@article{hanai2025photoinduced,
  title={Photoinduced non-reciprocal magnetism},
  author={Hanai, Ryo and Ootsuki, Daiki and Tazai, Rina},
  journal={Nature communications},
  volume={16},
  number={1},
  pages={8195},
  year={2025},
  publisher={Nature Publishing Group UK London},
  doi={https://doi.org/10.6084/m9.figshare.26334430},
  url={https://www.nature.com/articles/s41467-025-62707-9}
}

@article{guo2025field,
  title = {{Field Theory of Monitored Interacting Fermion Dynamics with Charge Conservation}},
  author = {Guo, Haoyu and Foster, Matthew S. and Jian, Chao-Ming and Ludwig, Andreas W. W.},
  year = 2025,
  journal = {Physical Review B},
  volume = {112},
  number = {6},
  pages = {064304},
  issn = {2469-9950, 2469-9969},
  doi = {10.1103/gdxd-pw8v},
  langid = {english}
}

@article{vansaders2026measurement,
    author = {VanSaders, Bryan and Fruchart, Michel and Vitelli, Vincenzo},
    title = {Measurement-induced phase transitions in informational active matter},
    journal = {PNAS Nexus},
    volume = {5},
    number = {4},
    pages = {pgag077},
    year = {2026},
    issn = {2752-6542},
    doi = {10.1093/pnasnexus/pgag077},
    url = {https://doi.org/10.1093/pnasnexus/pgag077}
}

@article{lumia2024measurement,
  title = {{Measurement-induced transitions beyond Gaussianity: A single particle description}},
  author = {Lumia, Luca and Tirrito, Emanuele and Fazio, Rosario and Collura, Mario},
  journal = {Phys. Rev. Res.},
  volume = {6},
  issue = {2},
  pages = {023176},
  numpages = {10},
  year = {2024},
  month = {May},
  publisher = {American Physical Society},
  doi = {10.1103/PhysRevResearch.6.023176},
  url = {https://link.aps.org/doi/10.1103/PhysRevResearch.6.023176}
}

@article{marchetti2013hydrodynamics,
  title = {Hydrodynamics of soft active matter},
  author = {Marchetti, M. C. and Joanny, J. F. and Ramaswamy, S. and Liverpool, T. B. and Prost, J. and Rao, Madan and Simha, R. Aditi},
  journal = {Rev. Mod. Phys.},
  volume = {85},
  issue = {3},
  pages = {1143--1189},
  numpages = {0},
  year = {2013},
  month = {Jul},
  publisher = {American Physical Society},
  doi = {10.1103/RevModPhys.85.1143},
  url = {https://link.aps.org/doi/10.1103/RevModPhys.85.1143}
}

@book{Gogolin1998, 
place={Cambridge}, 
title={{Bosonization and Strongly Correlated Systems}}, 
publisher={Cambridge University Press}, 
author={A. O. Gogolin and A. A. Nersesyan and A. M. Tsvelik},
year={1998}
}

@Article{Hoke2023,
author={Hoke, J. C.
and Ippoliti, M.
and Rosenberg, E.
and Abanin, D.
and Acharya, R.
and Andersen, T. I.
and Ansmann, M.
and Arute, F.
and Arya, K.
and Asfaw, A.
and {et al. (Google Quantum AI)} },
title={Measurement-induced entanglement and teleportation on a noisy quantum processor},
journal={Nature},
year={2023},
month={Oct},
day={01},
volume={622},
number={7983},
pages={481-486},
doi={10.1038/s41586-023-06505-7},
url={https://doi.org/10.1038/s41586-023-06505-7}
}

@book{Nazarov2009,
  title={{Quantum Transport: Introduction to Nanoscience}},
  author={Yuli V. Nazarov and Yaroslav M. Blanter},
  year={2009},
  publisher={Cambridge University Press, Cambridge, UK}
}

@book{nielsen2010quantum,
  title={{Quantum computation and quantum information}},
  author={Nielsen, Michael A and Chuang, Isaac L},
  year={2010},
  publisher={Cambridge University Press, Cambridge, UK}
}

@Article{Koh2023,
author={Koh, Jin Ming
and Sun, Shi-Ning
and Motta, Mario
and Minnich, Austin J.},
title={Measurement-induced entanglement phase transition on a superconducting quantum processor with mid-circuit readout},
journal={Nat. Phys.},
year={2023},
month={Sep},
day={01},
volume={19},
number={9},
pages={1314-1319},
doi={10.1038/s41567-023-02076-6},
url={https://doi.org/10.1038/s41567-023-02076-6}
}

@article{bechinger2016active,
	title = {Active {Particles} in {Complex} and {Crowded} {Environments}},
	volume = {88},
	url = {https://link.aps.org/doi/10.1103/RevModPhys.88.045006},
	doi = {10.1103/RevModPhys.88.045006}, 
	number = {4}, 
	journal = {Rev. Mod. Phys.},
	author = {Bechinger, Clemens and Di Leonardo, Roberto and Löwen, Hartmut and Reichhardt, Charles and Volpe, Giorgio and Volpe, Giovanni},
	month = nov,
	year = {2016},
	pages = {045006}
}

@article{ramaswamy2017active,
doi = {10.1088/1742-5468/aa6bc5},
url = {https://doi.org/10.1088/1742-5468/aa6bc5},
year = {2017},
volume = {2017},
number = {5},
pages = {054002},
author = {Ramaswamy, Sriram},
title = {Active matter},
journal = {J. Stat. Mech.: Theo. Exp.}
}

@article{shankar2022topological,
	title = {Topological active matter},
	volume = {4},
	url = {https://www.nature.com/articles/s42254-022-00445-3},
	doi = {10.1038/s42254-022-00445-3},
	number = {6},
	journal = {Nat. Rev. Phys.},
	author = {Shankar, Suraj and Souslov, Anton and Bowick, Mark J. and Marchetti, M. Cristina and Vitelli, Vincenzo},
	year = {2022},
	pages = {380--398}
}

@article{vicsek1995novel,
  title = {{Novel Type of Phase Transition in a System of Self-Driven Particles}},
  author = {Vicsek, Tam\'as and Czir\'ok, Andr\'as and Ben-Jacob, Eshel and Cohen, Inon and Shochet, Ofer},
  journal = {Phys. Rev. Lett.},
  volume = {75},
  issue = {6},
  pages = {1226--1229},
  numpages = {0},
  year = {1995},
  month = {Aug},
  publisher = {American Physical Society},
  doi = {10.1103/PhysRevLett.75.1226},
  url = {https://link.aps.org/doi/10.1103/PhysRevLett.75.1226}
}

@article{solon2013revisiting,
	title = {Revisiting the {Flocking} {Transition} {Using} {Active} {Spins}},
	volume = {111},
	copyright = {http://link.aps.org/licenses/aps-default-license},
	issn = {0031-9007, 1079-7114},
	url = {https://link.aps.org/doi/10.1103/PhysRevLett.111.078101},
	doi = {10.1103/PhysRevLett.111.078101},
	number = {7},
	urldate = {2024-11-11},
	journal = {Phys. Rev. Lett.},
	author = {Solon, A. P. and Tailleur, J.},
	month = aug,
	year = {2013},
	pages = {078101}
}

@article{chate2020dry,
	title = {Dry {Aligning} {Dilute} {Active} {Matter}},
	volume = {11},
	issn = {1947-5454, 1947-5462},
	url = {https://www.annualreviews.org/content/journals/10.1146/annurev-conmatphys-031119-050752},
	doi = {10.1146/annurev-conmatphys-031119-050752},
	number = {Volume 11, 2020},
	urldate = {2025-12-15},
	journal = {Annu. Rev. Condens. Matter Phys.},
	author = {Chaté, Hugues},
	month = mar,
	year = {2020},
	pages = {189--212}
}

@article{benvegnen2022flocking,
	title = {Flocking in one dimension: {Asters} and reversals},
	volume = {106},
	issn = {2470-0045, 2470-0053},
	shorttitle = {Flocking in one dimension},
	url = {https://link.aps.org/doi/10.1103/PhysRevE.106.054608},
	doi = {10.1103/PhysRevE.106.054608},
	number = {5},
	urldate = {2025-12-15},
	journal = {Phys. Rev. E},
	author = {Benvegnen, Brieuc and Chaté, Hugues and Krapivsky, Pavel L. and Tailleur, Julien and Solon, Alexandre},
	month = nov,
	year = {2022},
	pages = {054608}
}

@article{toner1995long,
  title = {{Long-Range Order in a Two-Dimensional Dynamical $\mathrm{XY}$ Model: How Birds Fly Together}},
  author = {Toner, John and Tu, Yuhai},
  journal = {Phys. Rev. Lett.},
  volume = {75},
  issue = {23},
  pages = {4326--4329},
  numpages = {0},
  year = {1995},
  month = {Dec},
  publisher = {American Physical Society},
  doi = {10.1103/PhysRevLett.75.4326},
  url = {https://link.aps.org/doi/10.1103/PhysRevLett.75.4326}
}

@article{solon2015phase,
	title = {From {Phase} to {Microphase} {Separation} in {Flocking} {Models}: {The} {Essential} {Role} of {Nonequilibrium} {Fluctuations}},
	volume = {114},
	copyright = {http://link.aps.org/licenses/aps-default-license},
	issn = {0031-9007, 1079-7114},
	shorttitle = {From {Phase} to {Microphase} {Separation} in {Flocking} {Models}},
	url = {https://link.aps.org/doi/10.1103/PhysRevLett.114.068101},
	doi = {10.1103/PhysRevLett.114.068101},
	number = {6},
	urldate = {2025-12-15},
	journal = {Phys. Rev. Lett.},
	author = {Solon, Alexandre P. and Chaté, Hugues and Tailleur, Julien},
	month = feb,
	year = {2015},
	pages = {068101}
}

@article{vrugt2025exactlyactivematter,
  title = {{Colloquium: What do we mean by `active matter'?}},
  author = {te Vrugt, Michael and Liebchen, Benno and Cates, Michael E.},
  journal = {Rev. Mod. Phys.},
  volume = {98},
  issue = {3},
  pages = {031001},
  numpages = {30},
  year = {2026},
  month = {Jul},
  publisher = {American Physical Society},
  doi = {10.1103/wd4f-q7kv},
  url = {https://link.aps.org/doi/10.1103/wd4f-q7kv}
}

@article{yamagishi_proposal_2024,
	title = {Proposal of a quantum version of active particles via a nonunitary quantum walk},
	volume = {14},
	copyright = {2024 The Author(s)},
	issn = {2045-2322},
	url = {https://www.nature.com/articles/s41598-024-78986-z},
	doi = {10.1038/s41598-024-78986-z},
	number = {1},
	urldate = {2025-12-16},
	journal = {Sci. Rep.},
	author = {Yamagishi, Manami and Hatano, Naomichi and Obuse, Hideaki},
	month = nov,
	year = {2024},
	pages = {28648}
}

@article{antonov2025engineering,
	title = {Engineering active motion in quantum matter},
	volume = {7},
	issn = {2643-1564},
	url = {https://link.aps.org/doi/10.1103/z3gm-32jn},
	doi = {10.1103/z3gm-32jn},
	number = {3},
	urldate = {2025-12-16},
	journal = {Phys. Rev. Res.},
	author = {Antonov, Alexander P. and Zheng, Yuanjian and Liebchen, Benno and Löwen, Hartmut},
	month = jul,
	year = {2025},
	pages = {033008}
}

@article{penner2025heat,
	title = {Heat-to-motion conversion for quantum active matter},
	volume = {112},
	issn = {2469-9950, 2469-9969},
	url = {https://link.aps.org/doi/10.1103/r6tm-nx19},
	doi = {10.1103/r6tm-nx19},
	number = {18},
	urldate = {2025-12-16},
	journal = {Phys. Rev. B},
	author = {Penner, Alexander-Georg and Viotti, Ludmila and Fazio, Rosario and Arrachea, Liliana and von Oppen, Felix},
	month = nov,
	year = {2025},
	pages = {L180303}
}

@misc{antonov2025modeling,
	title = {Modeling dissipation in quantum active matter},
	url = {http://arxiv.org/abs/2511.21502},
	doi = {10.48550/arXiv.2511.21502},
	publisher = {arXiv},
	author = {Antonov, Alexander P. and Lee, Sangyun and Liebchen, Benno and Löwen, Hartmut and Melles, Jannis and Morigi, Giovanna and Tuchkov, Yehor and te Vrugt, Michael},
	month = nov,
	year = {2025},
	note = {arXiv:2511.21502}
}

@misc{kundu2025run,
  title={Run-and-Tumble Dynamics and Zeno--Anti-Zeno Transition in Biased Quantum Trajectories},
  author={Kundu, Aritra},
  eprint={2512.20519},
  year={2025},
  archivePrefix={arXiv},
  url={https://arxiv.org/abs/2512.20519}, 
}

@article{khasseh2023active,
  title = {{Active Quantum Flocks}},
  author = {Khasseh, Reyhaneh and Wald, Sascha and Moessner, Roderich and Weber, Christoph A. and Heyl, Markus},
  journal = {Phys. Rev. Lett.},
  volume = {135},
  issue = {24},
  pages = {248302},
  numpages = {8},
  year = {2025},
  month = {Dec},
  publisher = {American Physical Society},
  doi = {10.1103/rd46-hr3q},
  url = {https://link.aps.org/doi/10.1103/rd46-hr3q}
}

@article{adachi2022activity,
	title = {Activity-induced phase transition in a quantum many-body system},
	volume = {4},
	issn = {2643-1564},
	url = {https://link.aps.org/doi/10.1103/PhysRevResearch.4.013194},
	doi = {10.1103/PhysRevResearch.4.013194},
	urldate = {2024-11-09},
	journal = {Phys. Rev. Res.},
	author = {Adachi, Kyosuke and Takasan, Kazuaki and Kawaguchi, Kyogo},
	month = mar,
	year = {2022},
	pages = {013194}
}

@article{takasan2024activity,
	title = {Activity-induced ferromagnetism in one-dimensional quantum many-body systems},
	volume = {6},
	issn = {2643-1564},
	url = {https://link.aps.org/doi/10.1103/PhysRevResearch.6.023096},
	doi = {10.1103/PhysRevResearch.6.023096},
	urldate = {2024-11-09},
	journal = {Phys. Rev. Res.},
	author = {Takasan, Kazuaki and Adachi, Kyosuke and Kawaguchi, Kyogo},
	month = apr,
	year = {2024},
	pages = {023096}
}

@article{skinner2019measurement,
	title = {Measurement-{Induced} {Phase} {Transitions} in the {Dynamics} of {Entanglement}},
	volume = {9},
	url = {https://link.aps.org/doi/10.1103/PhysRevX.9.031009},
	doi = {10.1103/PhysRevX.9.031009},
	number = {3},
	urldate = {2020-07-09},
	journal = {Phys. Rev. X},
	author = {Skinner, Brian and Ruhman, Jonathan and Nahum, Adam},
	month = jul,
	year = {2019},
	pages = {031009}
}

@article{li2018quantum,
	title = {Quantum {Zeno} effect and the many-body entanglement transition},
	volume = {98},
	url = {https://link.aps.org/doi/10.1103/PhysRevB.98.205136},
	doi = {10.1103/PhysRevB.98.205136},
	number = {20},
	urldate = {2020-07-09},
	journal = {Phys. Rev. B},
	author = {Li, Yaodong and Chen, Xiao and Fisher, Matthew P. A.},
	month = nov,
	year = {2018},
	pages = {205136}
}

@article{gullans2020dynamical,
  title = {{Dynamical Purification Phase Transition Induced by Quantum Measurements}},
  author = {Gullans, Michael J. and Huse, David A.},
  journal = {Phys. Rev. X},
  volume = {10},
  issue = {4},
  pages = {041020},
  numpages = {28},
  year = {2020},
  month = {Oct},
  publisher = {American Physical Society},
  doi = {10.1103/PhysRevX.10.041020},
  url = {https://link.aps.org/doi/10.1103/PhysRevX.10.041020}
}

@article{choi2019quantum,
  title = {{Quantum Error Correction in Scrambling Dynamics and Measurement-Induced Phase Transition}},
  author = {Choi, Soonwon and Bao, Yimu and Qi, Xiao-Liang and Altman, Ehud},
  journal = {Phys. Rev. Lett.},
  volume = {125},
  issue = {3},
  pages = {030505},
  numpages = {6},
  year = {2020},
  month = {Jul},
  doi = {10.1103/PhysRevLett.125.030505},
  url = {https://link.aps.org/doi/10.1103/PhysRevLett.125.030505}
}

@article{jian2020measurement,
  title = {Measurement-induced criticality in random quantum circuits},
  author = {Jian, Chao-Ming and You, Yi-Zhuang and Vasseur, Romain and Ludwig, Andreas W. W.},
  journal = {Phys. Rev. B},
  volume = {101},
  issue = {10},
  pages = {104302},
  numpages = {11},
  year = {2020},
  month = {Mar},
  publisher = {American Physical Society},
  doi = {10.1103/PhysRevB.101.104302},
  url = {https://link.aps.org/doi/10.1103/PhysRevB.101.104302}
}

@article{bao2020theory,
  title = {Theory of the phase transition in random unitary circuits with measurements},
  author = {Bao, Yimu and Choi, Soonwon and Altman, Ehud},
  journal = {Phys. Rev. B},
  volume = {101},
  issue = {10},
  pages = {104301},
  numpages = {26},
  year = {2020},
  month = {Mar},
  doi = {10.1103/PhysRevB.101.104301},
  url = {https://link.aps.org/doi/10.1103/PhysRevB.101.104301}
}

@article{barratt2022field,
  title = {Field {{Theory}} of {{Charge Sharpening}} in {{Symmetric Monitored Quantum Circuits}}},
  author = {Barratt, Fergus and Agrawal, Utkarsh and Gopalakrishnan, Sarang and Huse, David A. and Vasseur, Romain and Potter, Andrew C.},
  year = 2022,
  journal = {Physical Review Letters},
  volume = {129},
  number = {12},
  pages = {120604},
  publisher = {American Physical Society},
  doi = {10.1103/PhysRevLett.129.120604}
}

@article{cao2019entanglement,
	title = {Entanglement in a fermion chain under continuous monitoring},
	volume = {7},
	issn = {2542-4653},
	url = {https://scipost.org/10.21468/SciPostPhys.7.2.024},
	doi = {10.21468/SciPostPhys.7.2.024},
	number = {2},
	urldate = {2025-02-05},
	journal = {SciPost Phys.},
	author = {Cao, Xiangyu and Tilloy, Antoine and De Luca, Andrea},
	month = aug,
	year = {2019},
	pages = {024}
}

@article{alberton2021entanglement,
	title = {Entanglement {Transition} in a {Monitored} {Free}-{Fermion} {Chain}: {From} {Extended} {Criticality} to {Area} {Law}},
	volume = {126},
	shorttitle = {Entanglement {Transition} in a {Monitored} {Free}-{Fermion} {Chain}},
	url = {https://link.aps.org/doi/10.1103/PhysRevLett.126.170602},
	doi = {10.1103/PhysRevLett.126.170602},
	journal = {Phys. Rev. Lett.},
	author = {Alberton, O. and Buchhold, M. and Diehl, S.},
	month = apr,
	year = {2021},
	pages = {170602}
}

@article{buchhold2021effective,
  title = {{Effective Theory for the Measurement-Induced Phase Transition of Dirac Fermions}},
  author = {Buchhold, M. and Minoguchi, Y. and Altland, A. and Diehl, S.},
  journal = {Phys. Rev. X},
  volume = {11},
  issue = {4},
  pages = {041004},
  numpages = {35},
  year = {2021},
  month = {Oct},
  doi = {10.1103/PhysRevX.11.041004},
  url = {https://link.aps.org/doi/10.1103/PhysRevX.11.041004}
}

@article{poboiko2023theory,
	title = {Theory of {Free} {Fermions} under {Random} {Projective} {Measurements}},
	volume = {13},
	url = {https://link.aps.org/doi/10.1103/PhysRevX.13.041046},
	doi = {10.1103/PhysRevX.13.041046},
	journal = {Phys. Rev. X},
	author = {Poboiko, Igor and Pöpperl, Paul and Gornyi, Igor V. and Mirlin, Alexander D.},
	month = dec,
	year = {2023},
	pages = {041046}
}

@article{fava2023nonlinear,
  title = {{Nonlinear Sigma Models for Monitored Dynamics of Free Fermions}},
  author = {Fava, Michele and Piroli, Lorenzo and Swann, Tobias and Bernard, Denis and Nahum, Adam},
  journal = {Phys. Rev. X},
  volume = {13},
  issue = {4},
  pages = {041045},
  numpages = {33},
  year = {2023},
  month = {Dec},
  doi = {10.1103/PhysRevX.13.041045},
  url = {https://link.aps.org/doi/10.1103/PhysRevX.13.041045}
}

@article{poboiko2025measurement,
	title = {Measurement-induced transitions for interacting fermions},
	volume = {111},
	url = {https://link.aps.org/doi/10.1103/PhysRevB.111.024204},
	doi = {10.1103/PhysRevB.111.024204},
	journal = {Phys. Rev. B},
	author = {Poboiko, Igor and Pöpperl, Paul and Gornyi, Igor V. and Mirlin, Alexander D.},
	month = jan,
	year = {2025},
	pages = {024204}
}

@article{mueller2025monitored,
  title = {{Monitored interacting Dirac fermions}},
  author = {M\"uller, T. and Buchhold, M. and Diehl, S.},
  journal = {Phys. Rev. B},
  volume = {111},
  issue = {17},
  pages = {174201},
  numpages = {23},
  year = {2025},
  month = {May},
  publisher = {American Physical Society},
  doi = {10.1103/PhysRevB.111.174201},
  url = {https://link.aps.org/doi/10.1103/PhysRevB.111.174201}
}

@article{Fisher2023,
   author = "Fisher, Matthew P.A. and Khemani, Vedika and Nahum, Adam and Vijay, Sagar",
   title = {{Random Quantum Circuits}}, 
   journal= "Annu. Rev. Condens. Matter Phys.",
   year = "2023",
   volume = "14",
   pages = "335-379",
   doi = "https://doi.org/10.1146/annurev-conmatphys-031720-030658",
   url = "https://www.annualreviews.org/content/journals/10.1146/annurev-conmatphys-031720-030658"
  }

@book{wiseman2009quantum,
  title={Quantum measurement and control},
  author={Wiseman, Howard M and Milburn, Gerard J},
  year={2009},
  publisher={Cambridge University Press}
}

@article{buchhold2015nonequilibrium,
	title = {Nonequilibrium universality in the heating dynamics of interacting {Luttinger} liquids},
	volume = {92},
	copyright = {http://link.aps.org/licenses/aps-default-license},
	issn = {1050-2947, 1094-1622},
	url = {https://link.aps.org/doi/10.1103/PhysRevA.92.013603},
	doi = {10.1103/PhysRevA.92.013603},
	urldate = {2025-12-13},
	journal = {Phys. Rev. A},
	author = {Buchhold, Michael and Diehl, Sebastian},
	month = jul,
	year = {2015},
	pages = {013603}
}

@article{egger_scaling_1998,
	title = {Scaling and criticality of the {Kondo} effect in a {Luttinger} liquid},
	volume = {57},
	url = {https://link.aps.org/doi/10.1103/PhysRevB.57.10620},
	doi = {10.1103/PhysRevB.57.10620},
	journal = {Phys. Rev. B},
	author = {Egger, Reinhold and Komnik, Andrei},
	month = may,
	year = {1998},
	pages = {10620--10629}, 
}

@article{fazio2025many,
	title = {Many-body open quantum systems},
	issn = {2590-1990},
	url = {https://scipost.org/SciPostPhysLectNotes.99},
	doi = {10.21468/SciPostPhysLectNotes.99},
	journal = {SciPost Phys. Lect. Notes},
	author = {Fazio, Rosario and Keeling, Jonathan and Mazza, Leonardo and Schirò, Marco},
	month = aug,
	year = {2025},
	pages = {99}
}

@article{hardt2025propelling,
	title = {Propelling ferrimagnetic domain walls by dynamical frustration},
	volume = {16},
	url = {https://www.nature.com/articles/s41467-025-58920-1},
	doi = {10.1038/s41467-025-58920-1},
	journal = {Nat. Commun.},
	author = {Hardt, Dennis and Doostani, Reza and Diehl, Sebastian and del Ser, Nina and Rosch, Achim},
	month = apr,
	year = {2025},
	pages = {3817}
}

@article{delser2023skyrmion,
	title = {Skyrmion jellyfish in driven chiral magnets},
	volume = {15},
	issn = {2542-4653},
	url = {https://scipost.org/10.21468/SciPostPhys.15.2.065},
	doi = {10.21468/SciPostPhys.15.2.065},
	journal = {SciPost Phys.},
	author = {Del Ser, Nina and Lohani, Vivek},
	month = aug,
	year = {2023},
	pages = {065}
}

@article{egger1996,
  title = {{RKKY interaction and Kondo screening cloud for strongly correlated electrons}},
  author = {Egger, Reinhold and Schoeller, Herbert},
  journal = {Phys. Rev. B},
  volume = {54},
  issue = {23},
  pages = {16337--16340},
  numpages = {0},
  year = {1996},
  month = {Dec},
  publisher = {American Physical Society},
  doi = {10.1103/PhysRevB.54.16337},
  url = {https://link.aps.org/doi/10.1103/PhysRevB.54.16337}
}

@article{calabrese2004entanglement,
doi = {10.1088/1742-5468/2004/06/P06002},
url = {https://doi.org/10.1088/1742-5468/2004/06/P06002},
year = {2004},
month = {jun},
volume = {2004},
number = {06},
pages = {P06002},
author = {Pasquale Calabrese and John Cardy},
title = {Entanglement entropy and quantum field theory},
journal = {J. Stat. Mech.: Theo. Exp.}
}

@article{dehghani2023neuralnetwork,
  title = {{Neural-Network Decoders for Measurement Induced Phase Transitions}},
  author = {Dehghani, Hossein and Lavasani, Ali and Hafezi, Mohammad and Gullans, Michael J.},
  year = 2023,
  journal = {Nat. Commun.},
  volume = {14},
  number = {1},
  pages = {2918},
  issn = {2041-1723},
  doi = {10.1038/s41467-023-37902-1},
  langid = {english}
}

@article{feng2026postselectionfree,
  title = {{Postselection-Free Experimental Observation of the Measurement-Induced Phase Transition in Circuits with Universal Gates}},
  author = {Feng, Xiaozhou and C{\^o}t{\'e}, Jeremy and Kourtis, Stefanos and Skinner, Brian},
  year = 2026,
  journal = {Commun. Phys.},
  publisher = {Nature Publishing Group},
  issn = {2399-3650},
  volume = {5},
  pages = {110},
  doi = {10.1038/s42005-025-02443-0}
}

@article{garratt2024probing,
  title = {Probing {{Postmeasurement Entanglement}} without {{Postselection}}},
  author = {Garratt, Samuel J. and Altman, Ehud},
  year = 2024,
  journal = {PRX Quantum},
  volume = {5},
  number = {3},
  pages = {030311},
  publisher = {American Physical Society},
  doi = {10.1103/PRXQuantum.5.030311}
}

@article{ippoliti2021postselectionfree,
  title = {Postselection-{{Free Entanglement Dynamics}} via {{Spacetime Duality}}},
  author = {Ippoliti, Matteo and Khemani, Vedika},
  year = 2021,
  journal = {Phys. Rev. Lett.},
  volume = {126},
  number = {6},
  pages = {060501},
  publisher = {American Physical Society},
  doi = {10.1103/PhysRevLett.126.060501}
}

@Misc{SI,
  note = {{See Supplementary Material at [URL] for details of monitoring protocol and the analytical calculations for the bosonized model.}}
}

@article{mcginley2024postselectionfree,
  title = {Postselection-{{Free Learning}} of {{Measurement-Induced Quantum Dynamics}}},
  author = {McGinley, Max},
  year = 2024,
  journal = {PRX Quantum},
  volume = {5},
  number = {2},
  pages = {020347},
  publisher = {American Physical Society},
  doi = {10.1103/PRXQuantum.5.020347}
}

@article{noel2022measurementinduced,
  title = {{Measurement-Induced Quantum Phases Realized in a Trapped-Ion Quantum Computer}},
  author = {Noel, Crystal and Niroula, Pradeep and Zhu, Daiwei and Risinger, Andrew and Egan, Laird and Biswas, Debopriyo and Cetina, Marko and Gorshkov, Alexey V. and Gullans, Michael J. and Huse, David A. and Monroe, Christopher},
  year = 2022,
  journal = {Nat. Phys.},
  volume = {18},
  number = {7},
  pages = {760--764},
  publisher = {Nature Publishing Group},
  issn = {1745-2481},
  doi = {10.1038/s41567-022-01619-7},
  copyright = {2022 The Author(s), under exclusive licence to Springer Nature Limited},
  langid = {english}
}

@article{antonic2025motion,
  title = {Motion from Measurement: {{The}} Role of Symmetry of Quantum Measurements},
  shorttitle = {Motion from Measurement},
  author = {Antoni{\'c}, Luka and Kafri, Yariv and Podolsky, Daniel and Turner, Ari M.},
  year = 2025,
  journal = {Physical Review B},
  volume = {111},
  number = {22},
  pages = {224305},
  publisher = {American Physical Society},
  doi = {10.1103/PhysRevB.111.224305}
}

@article{hovhannisyan2019quantum,
  title = {{Quantum Current in Dissipative Systems}},
  author = {Hovhannisyan, Karen V. and Imparato, Alberto},
  year = 2019,
  journal = {New Journal of Physics},
  volume = {21},
  number = {5},
  pages = {052001},
  publisher = {IOP Publishing},
  issn = {1367-2630},
  doi = {10.1088/1367-2630/ab1731},
  langid = {english}
}

@article{popperl2023measurements,
  title = {Measurements on an {{Anderson}} Chain},
  author = {P{\"o}pperl, Paul and Gornyi, Igor V. and Gefen, Yuval},
  year = 2023,
  journal = {Physical Review B},
  volume = {107},
  number = {17},
  pages = {174203},
  issn = {2469-9950, 2469-9969},
  doi = {10.1103/PhysRevB.107.174203},
  langid = {english}
}

@article{suzuki2025quantum,
  title = {{Quantum Complexity Phase Transitions in Monitored Random Circuits}},
  author = {Suzuki, Ryotaro and Haferkamp, Jonas and Eisert, Jens and Faist, Philippe},
  year = 2025,
  journal = {Quantum},
  volume = {9},
  pages = {1627},
  publisher = {Verein zur F\"orderung des Open Access Publizierens in den Quantenwissenschaften},
  doi = {10.22331/q-2025-02-10-1627},
  langid = {british}
}

@article{gross2018qubit,
doi = {10.1088/2058-9565/aaa39f},
url = {https://doi.org/10.1088/2058-9565/aaa39f},
year = {2018},
month = {feb},
publisher = {IOP Publishing},
volume = {3},
number = {2},
pages = {024005},
author = {Gross, Jonathan A and Caves, Carlton M and Milburn, Gerard J and Combes, Joshua},
title = {Qubit models of weak continuous measurements: markovian conditional and open-system dynamics},
journal = {Quantum Science and Technology}
}
\onecolumngrid
\newpage

\appendix
\section{Monitoring protocol}\label{sm:directed_hopping_protocol}

\begin{figure}[t]
\begin{center}
    \includegraphics[width=0.9\textwidth]{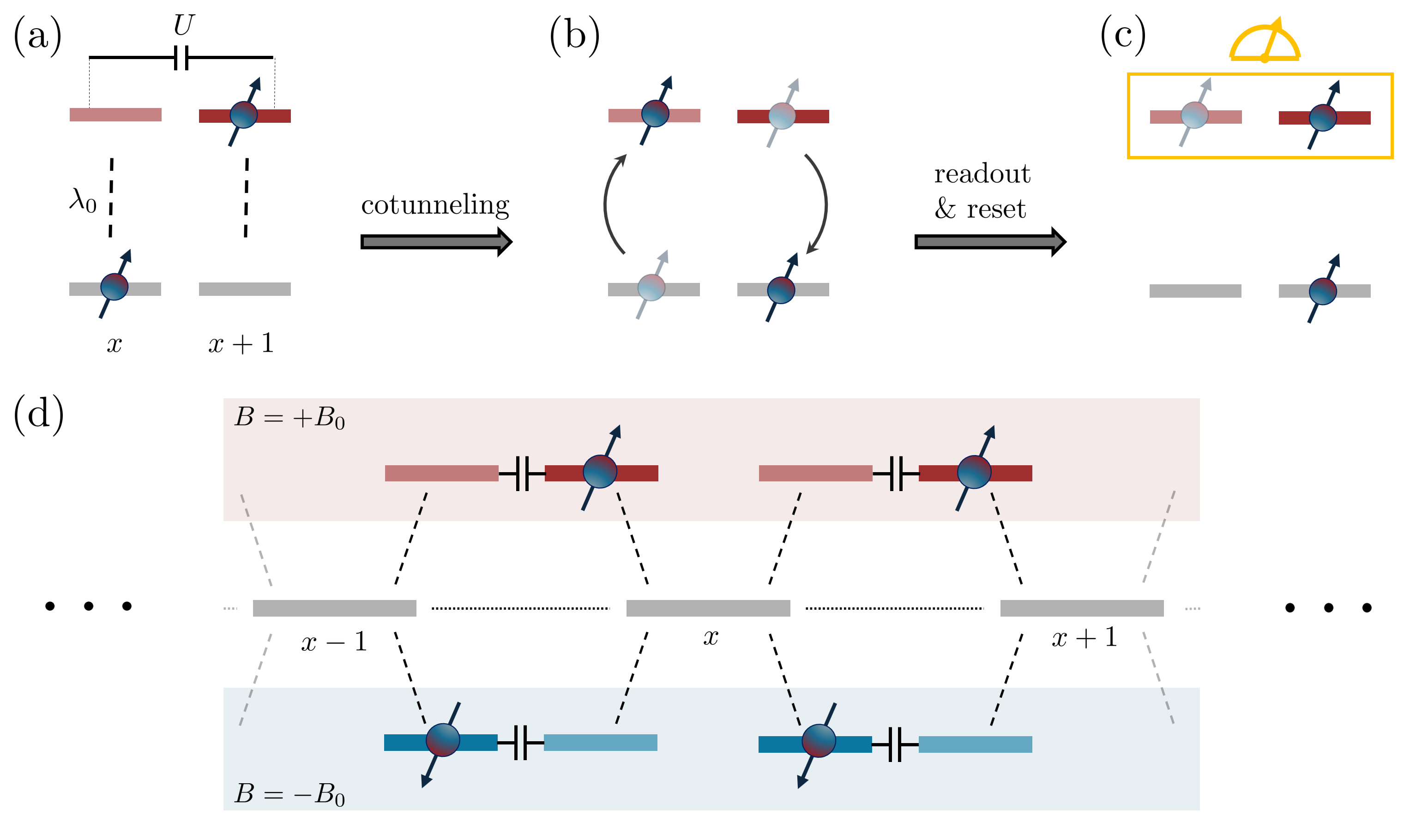}
    \caption{Schematic illustration of the setup and protocol realizing the jump operators in Eq.~\eqref{eq:lattice_jump_operators}. Panels (a)--(c) show how directional hopping of a spin-up fermion between neighboring sites $x$ and $x+1$ (gray levels) is implemented by tunnel-coupling (dashed lines, amplitude $\lambda_0$) each site to one dot of an auxiliary double dot (red levels). The auxiliary double dot is subject to a strongly repulsive interdot Coulomb repulsion $U$ and is operated in the single-occupancy regime. The auxiliary double dot is initialized with the fermion occupying the right dot, see panel (a). If a fermion occupies site $x$ of the chain it has amplitude $\lambda_\textrm{cot}$ to undergo a cotunneling process from $x$ to $x+1$ via the auxiliary double dot, leaving the latter in a state where the fermion now occupies the left dot, see panel (b). This change is detected by a projective measurement of the charge state of the auxiliary double dot which is performed after a time $\delta t$: finding the fermion on the left dot heralds the cotunneling process. Such projective measurements lead to a stochastic Schrödinger equation of the form Eq.~\eqref{eq:sse_poisson} with quantum jump noise. In order to realize a stochastic Schrödinger equation with Langevin noise such as Eq. \eqref{eq:sse}, one instead measures the double dot in a basis that is only weakly correlated with the occurrence of a cotunneling event, see discussion preceding Eq.~\eqref{eq:sse_gaussian}. 
    The auxiliary double dot is subsequently reset to its initial state, see orange box in panel (c), and the protocol is repeated. Finally, to implement directed hopping processes for both spin-up and spin-down fermions along the full chain, two auxiliary double dots are introduced for each link (red and blue levels, respectively), with different initial configurations, see panel (d): for the red double dots the fermion is initially on the right,  implementing rightward jumps [as shown in panels (a)-(c)], while for the blue double dots the fermion is initially on the left, implementing leftward jumps instead. Zeeman or magnetic exchange fields $B=\pm B_0$ (blue and red shaded regions) suppress the unwanted processes, blocking spin-down fermions from undergoing rightward jumps and spin-up fermions from undergoing leftward jumps. 
    For details, see Sec.~\ref{sm:directed_hopping_protocol} of the SM.
    }
    \label{fig:fig_S1}
\end{center} 
\end{figure}

We discuss how monitoring of the jump operators in Eq.~\eqref{eq:lattice_jump_operators}
can be implemented using a chain of spinful single-level quantum dots described by the fermion operators $c_{x,\sigma}$. The basic idea is to introduce additional parallel tunneling paths between neighboring sites, one each for spin-up fermions and for spin-down fermions. To implement unidirectional hopping, transmission through these tunneling paths involves a cotunneling process via a tunnel-coupled auxiliary double quantum dot in the single-occupancy regime. The double dot is maintained in a state that allows this cotunneling process to occur only in one direction. Monitoring is implemented by frequent projective readout of the quantum dot, which contain information on whether a cotunneling event has occurred or not. The rate at which these hops occur as well as the time between measurements control the monitoring strength.  For a schematic illustration of the setup, see Fig.~\ref{fig:fig_S1}. In the following, we first describe this protocol in the simplest possible scenario: spinless fermions and a single pair of sites. We then show how to extend this to the spinful setup considered in the main text. 

\subsection{Minimal example}

We first consider a chain of spinless fermions $c_x$ and use weak measurements in order to monitor directed jumps on a single link, $L^{}_{x} = \sqrt{\gamma} c^\dagger_{x+1} c^{}_x$. To this end, we introduce two auxiliary detector dots which are described by spinful fermion operators $d^{}_{i\in\{1,2\},\sigma}$. The Hamiltonian describing these detector dots is given by \cite{Nazarov2009}
\begin{equation}
    H_{L_{x}} = \lambda_0  \pqty{d^\dagger_{1} c^{}_{x} +  c^\dagger_{x+1} d^{}_{2} + \textrm{h.c.}}  
    + U \pqty{n^d_{1} -\frac{1}{2}} \pqty{n^d_{2}  -\frac{1}{2}},  
\end{equation}
with occupation numbers $n^d_{i}= d^\dagger_{i}d^{}_{i}$. Site $x$ ($x+1$) is tunnel-coupled to the detector dot $1$ ($2$) with amplitude $\lambda_0 > 0$. The two dots are in the strong Coulomb repulsion regime $U \gg \lambda_0$ and subject to a chemical potential offset $-U/2$, such that the empty and doubly occupied states, $\ket{\emptyset}_d$ and $d_1^\dagger d_2^\dagger\ket{\emptyset}_d$, respectively, are pushed to high energies. Eliminating the high-energy states gives the cotunneling Hamiltonian 
\begin{equation}
    H_{L_{x}} \simeq  \lambda_\textrm{cot}  \pqty{  c^\dagger_{x+1} c^{}_{x} \otimes \ketbra{1}{0}_d +  c^\dagger_{x} c^{}_{x+1} \otimes  \ketbra{0}{1}_d },
\end{equation}
in terms of the singly-occupied states $\ket{0}_d = d_2^\dagger\ket{\emptyset}_d$ and $\ket{1}_d = d_1^\dagger\ket{\emptyset}_d$ and the cotunneling amplitude   $\lambda_\textrm{cot} \sim \lambda_0^2/U$ \cite{Nazarov2009}. To realize the desired jump operators, we employ the following protocol starting from the initial state, in which the auxiliary double dot is in the state $\ket{0}_d$ (i.e., the right dot is occupied):
\begin{enumerate} 
    \item Time-evolve the combined double-dot--chain system under $H_\textrm{total} = H + H_{L_x}$, where $H$ is the chain Hamiltonian, for some time $\delta t$ which is small compared to the inverse cotunneling amplitude as well as system energy scales. To leading order, the resulting state is
    \begin{equation}
        e^{-i( H + H_{L_x})\delta t} \ket{\psi}  \ket{0}_d \simeq \bqty{1-i \delta t \pqty{ H - \frac{1}{2} L_x^\dagger L^{}_x }  }\ket{\psi}  \ket{0}_d - i \sqrt{\delta t } L^{}_x \ket{\psi} \ket{1}_d ,
    \end{equation}
    where $\ket{\psi} $ is the state of the fermion chain and $L^{}_x = \sqrt{\gamma} c^\dagger_{x+1}c_x$ with $\gamma = \delta t\lambda_\textrm{cot}^2$. The term $\sim L_x^\dagger L^{}_x \delta t$ has to be kept to ensure proper normalization of the post-measurement state.  
    \item Projectively measure the auxiliary double dot in the charge basis. The two possible outcomes are $\ket{e=0}_d$, i.e., no cotunneling event has occurred during the period $\delta t$, and $\ket{e = 1}_d \equiv d_2^\dagger \ket{\emptyset}_d$, i.e., a cotunneling event has occurred. 
    \item Reset the state of the auxiliary double dot back to $\ket{0}_d$. 
    \item Repeat the above steps.
\end{enumerate} 
One round of this protocol can be described in terms of Kraus operators $\mathcal{K}_{e} = \bra{e}_d e^{-i\delta t [H + H_{L_{x}}]} \ket{0}_d$ \cite{nielsen2010quantum}. They are
\begin{equation}\label{eq:Kraus}
    \mathcal{K}^{}_{0} = 1-i \delta t \pqty{ H - \frac{1}{2} L_x^\dagger L_x },\quad \mathcal{K}^{}_{1} = \sqrt{\delta t } L_x .
\end{equation} 
The probability to obtain outcome $e$ is $\bra{\psi}  \mathcal{K}_{e}^\dagger \mathcal{K}^{}_{e}\ket{\psi}$. The (unnormalized) system state after $n$ rounds is $\prod^n_{i=1} \mathcal{K}_{e_i} \ket{\psi}$, where $e_i$ denotes the outcome of the $i^\textrm{th}$ measurement. One can readily show that in the continuum limit this protocol realizes a stochastic Schrödinger equation of the form \cite{wiseman2009quantum}
\begin{equation}\label{eq:sse_poisson}
    d\ket{\psi_c} = - i \bqty{ H - \frac{1}{2} \pqty{L_x^\dagger L_x - \bra{\psi}L_x^\dagger L^{}_x \ket{\psi}}} \ket{\psi_c}  dt + \bqty{ \frac{L_x}{\sqrt{\bra{\psi_c}L_x^\dagger L^{}_x \ket{\psi_c}}} - 1}  \ket{\psi_c}  dN, 
\end{equation}
where the Poisson increment satisfies $dN^2 = dN$ and $\overline{dN} = \bra{\psi_c} L_x^\dagger L^{}_x\ket{\psi_c} dt$, and we introduced the subscript $c$ to denote conditioning on the measurement record. 

While we do not expect this to be essential for the qualitative physics, it is convenient to work with a Gaussian stochastic Schrödinger equation for calculational purposes. In principle, one can realize a Gaussian stochastic Schrödinger equation such as Eq.~\eqref{eq:sse} by slightly augmenting the above protocol. Coarse graining Eq.~\eqref{eq:sse_poisson} in time results in a Gaussian stochastic Schrödinger equation only if the jump operator is almost proportional to the identity, as, e.g., in homodyne detection or weak charge sensing. This is not the case for our $L_x$. We thus follow a different approach based on measurements in the basis $\ket{\pm}_d = (\ket{0}_d \pm  \ket{1}_d)/\sqrt{2}$. Such measurements can be realized by applying a unitary rotation to the auxiliary double dot prior to the measurement step. The outcomes $\pm$ correspond to Kraus operators $\mathcal{K}_\pm$ and occur with probabilities ${\rm Pr}_\pm$ given by
\begin{equation}
    \mathcal{K}_\pm = \frac{1}{\sqrt{2}} \bqty{1 - i \delta t \pqty{ H - \frac{1}{2} L_x^\dagger L_x } \mp \sqrt{\delta t}  L_x },\quad \textrm{Pr}_\pm = \frac{1}{2} \pqty{1 \pm \sqrt{\delta t} \bra{\psi} ( L_x^\dagger + L_x) \ket{\psi}}, 
\end{equation}
respectively. Thus, both Kraus operators are close to the identity, allowing for a description in terms of Gaussian white noise \cite{gross2018qubit}. The resulting stochastic Schrödinger equation has the desired form, 
\begin{equation}\label{eq:sse_gaussian}
    i\frac{d}{dt}\ket{\psi_c} =\Bqty{-iH - \frac{1}{2}   \bqty{L^\dagger_{x} L^{}_{x} - 2W^{}_{x} L^{}_{x} + W^2_{x} }
    +  \xi_{x } ( L^{}_{x} - W^{}_{x}) } \ \ket{\psi_c},
\end{equation}
where $W_x = \frac{1}{2} \bra{\psi_c} (L^\dagger_x +  L^{}_x) \ket{\psi_c}$. Note that directionality enters the protocol by the choice of reset state. In our example, the double dot is reset such that a fermion occupies the second (or right) dot prior to each interaction with the chain. A cotunneling process then effectively moves a chain fermion to the right. Choosing to initialize in, and reset to, the state with a fermion occupying the first (or left) dot instead results in a leftward cotunneling process.

\subsection{Generalization and time scales}

The protocol can be straightforwardly generalized to simultaneous measurements of directed jump operators on all bonds and with opposite directionality for the two spin species. To this end, for each bond $x,x+1$ we introduce two double quantum dots, one each per spin species. This is depicted in Fig.~\ref{fig:fig_S1}(d), with double quantum dots for spin-up fermions shown in red, and those for spin-down fermions shown in blue. We introduce a strong Zeeman or magnetic exchange field $B = \pm B_0$ with $B_0 \gg \lambda_0$ that effectively blocks fermions with the unwanted spin from entering the double dots. These fields are represented by red and blue shading in Fig.~\ref{fig:fig_S1}(d). We keep the auxiliary double dots in the strong Coulomb-blockade regime such that always exactly one fermion occupies a given double dot, and cotunneling of an electron with the appropriate spin is the only low-energy process. Concretely, the low-energy states of the auxiliary double dots responsible for $\uparrow$-fermion jumps are the states $\ket{0}_{d,\uparrow} = d^\dagger_{2,\uparrow}\ket{\emptyset}_{d}, \ket{1}_{d,\uparrow} = d^\dagger_{1,\uparrow}\ket{\emptyset}_{d}$, and for $\downarrow$-fermion jumps they are $\ket{0}_{d,\downarrow} = d^\dagger_{1,\downarrow}\ket{\emptyset}_{d}, \ket{1}_{d,\downarrow} = d^\dagger_{2,\downarrow}\ket{\emptyset}_{d}$. Note that for $\uparrow$-fermions, the reset state $\ket{0}_{d,\uparrow}$ has the right dot occupied, while for $\downarrow$-fermions, the reset state $\ket{0}_{d,\downarrow}$ has the left dot occupied. This choice controls the directionality, ensuring that the corresponding processes move fermions of opposite spins in opposite directions on the chain. With this construction, essentially the same arguments apply as in the case of a single link for spinless fermions. Indeed, the order in which one performs the projective measurements of the auxiliary double dots does not matter since the noncommuting nature of the corresponding Kraus operators enters only at higher orders in $\delta t$. 

To conclude the discussion of a concrete practical implementation of our proposal for active quantum matter based on quantum measurements, we comment on the required hierarchy of energy and timescales. First, in order for the occupation of the high-energy states of the detector dots to remain purely virtual, we require the Coulomb energy $U$ and Zeeman fields $B_0$ to be large compared to $\lambda_0$ as well as compared to system energy scales $t_0,J_z$. We note that the scale $U$ can be tuned either electrostatically, i.e., by means of gates, or by changing the geometric size of the double dot system. Similarly, the tunnel couplings governing $\lambda_0,t_0$ and $J_z$ are also gate-tunable and thus in principle can be made as small as one pleases.  Second, the time required for projective readout and reset of the auxiliary double quantum dots should be very short compared to the cotunneling rate $\lambda^{-1}_\textrm{cot}$ as well as $t_0^{-1}$ and $J_z^{-1}$ such that the system does not appreciably evolve during the readout. Finally, the time $\delta t$ controls the strength of the measurement for a given cotunneling rate. It should be chosen sufficiently small compared to the cotunneling time $\lambda_\textrm{cot}^{-1}$ as well as the system time scales $t_0^{-1},J_z^{-1}$ such that expansion of the time-evolution operator remains valid.

\section{Bosonization conventions}

For convenience, we briefly summarize our bosonization conventions following Ref.~\cite{egger_scaling_1998}. First, we let $c_{x,\sigma}^{} \to \sqrt{a} \psi_\sigma(x)$. We set the lattice constant $a \equiv 1$, i.e., we measure lengths in units of $a$. The fermionic field $\psi_\sigma(x)$ is decomposed into left and right moving contributions near the Fermi points $\pm k_F$,
$\psi_\sigma(x) \to \sum_{p=\pm} \psi_{\sigma,p }(x).$
We bosonize the latter fields via
\begin{equation}
    \psi_{\sigma,p}(x) = \frac{\eta_{\sigma,p}}{\sqrt{2\pi}}       e^{ip k_F x-i\sqrt{\pi}\theta_\sigma(x) + i p \sqrt{\pi} \varphi_\sigma(x)  },
\end{equation}
where $\theta_\sigma = (\theta_c+\sigma\theta_s)/\sqrt{2}$ and $\varphi_\sigma = (\varphi_c+\sigma\varphi_s)/\sqrt{2}$.
We note that $\partial_x\varphi_\sigma$ is proportional to the density of spin-$\sigma$ particles (with respect to the average value), while $\partial_x\theta_\sigma$ is proportional to the current density \cite{Gogolin1998}.
Anticommutation relations for fermions are enforced by Klein factors which can be written as Majorana operators $\eta_{\sigma,p}$ satisfying $\eta_{\sigma,p}^2 = 1$. As we restrict attention to correlations of fermion bilinears (densities and/or currents), we may  parameterize their products by Pauli matrices $\beta_{1,2,3}$ acting on an auxiliary Hilbert space, 
\begin{equation}  
    \eta_{\sigma,p}\eta_{\sigma,-p} = ip\sigma \beta_3,\quad
    \eta_{\sigma,p}\eta_{-\sigma,p} = i\sigma \beta_2,\quad 
   \eta_{\sigma,p}\eta_{-\sigma,-p} = ip \beta_1.  \label{pauli}
\end{equation}
Applying these bosonization rules, we obtain the bosonized Hamiltonian in Eq.~\eqref{eq:bosonized_hamiltonian} and the linear piece of the bosonized jump operators in Eq.~\eqref{eq:bosonized_jump_operators_lin}.
Note that we dropped a constant contribution to $M_{0,\sigma}$ proportional to the overall charge density $\varrho_0$. Keeping this term gives rise to a contribution $\propto \int dx\, \partial_x \theta_s$ to the Hamiltonian which does not affect the dynamics. One furthermore finds the nonlinear piece in Eq.~\eqref{eq:bosonized_jump_operators_nonlin}, which should technically be multiplied by $\beta_3$. In the absence of spin-flip terms, one can simply fix $\beta_3 = 1$ since no other $\beta_i$ feature in the problem. Spin-flip terms lead to contributions $\propto \beta_2$, such that one in principle needs to keep the auxiliary Pauli structure. However, the steady state is independent of such terms as, upon tracing over the absolute replica sector, no dependence on spin-flip processes remains. This is explained in detail below.

\section{Replica master equation}

Consider a monitored system evolving under the stochastic Schrödinger equation \eqref{eq:sse}. The ensemble-averaged two-replica state $\rho_2(t)$ then evolves according to  
\begin{equation}\label{eq:2_replica_eom_1}
    \dot{\rho}_2 =\pqty{ \mathcal{L}^{(1)} +  \mathcal{L}^{(2)} }\rho_2  + \sum_\mu \overline{\mathcal{R}[\tilde{L}_\mu]\rho_c \otimes \mathcal{R}[\tilde{L}_\mu]\rho_c}.
\end{equation}
Here, $\rho_c = \ketbra{\psi_c}$ and we define the single-replica Liouvillian as 
$\mathcal{L} \rho = - i \bqty{H,\rho}  +  \sum_\mu \mathcal{D}[L_\mu] \rho.$
We also introduced the multi-index $\mu = (x,\sigma)$ and the shorthands $\mathcal{D}[L]\rho = L \rho L^\dagger - \frac{1}{2}\{L^\dagger L , \rho\}$ with the anticommutator $\{ \cdot, \cdot \}$, $\mathcal{R}[L]\rho = L \rho + (L\rho)^\dagger$, and $\tilde{L}\rho = L\rho - \tr(L\rho)\rho$. Consider the last term in Eq.~\eqref{eq:2_replica_eom_1}. It contains third- and fourth-order terms in the system state, which couple the problem to sectors with higher replica number. With $M = (L + L^\dagger)/2$,
we proceed by the approximate decoupling 
\begin{equation}\label{eq:decoupling}
    \overline{\tr\pqty{\mathcal{R}[L]\rho_c}  \rho_c \otimes  \rho_c  }
    \simeq  2  \ev*{M^{(i)}} \rho_2 , \quad
    \overline{ \tr^2\pqty{\mathcal{R}[L]\rho_c}  \rho_c \otimes  \rho_c} \simeq 4  \ev*{M^{(1)} M^{(2)}} \rho_2. 
\end{equation}
We emphasize that this decoupling step yields identical results as calculations based on a linear stochastic master equation employing the standard replica trick to connect to physical observables \cite{fava2023nonlinear,mueller2025monitored}. Expectation values that only depend on a single replica are simply those of the measurement-averaged state, $\ev*{M^{(i)}} = \overline{\tr(M \rho_c)\tr(\rho_c)} = \tr(M \rho_1)$.  These expectation values vanish at sufficiently late times 
due to the heating-up of $\rho_1$ and the tracelessness of $M$.
Enforcing conservation of $\tr (\rho_2)$, we obtain
\begin{equation}\label{eq:2_replica_eom_2}
     \dot{\rho}_2  = \sum_i  \mathcal{L}^{(i)}  \rho_2   +  \sum_{\mu}  \mathcal{R}[L_\mu^{(1)} ]\mathcal{R}[L_\mu^{(2)} ]\rho_2 - \mathcal{N}\rho_2, 
\end{equation}
where $\mathcal{N} = 4\sum_\mu  \ev*{M_\mu^{(1)} M_\mu^{(2)}}$. 

In the bosonized continuum framework, after introducing relative ($r$) and absolute ($a$) boson fields in replica space, Eq.~\eqref{eq:2_replica_eom_2} takes the form 
\begin{equation}\label{eq:2repl}
\dot{\rho}_2  = \left(\mathcal{L}^{(a)}_0  + \mathcal{L}^{(r)}_0 + \delta \mathcal{L}\right)\rho_2.
\end{equation}
Here, $\mathcal{L}^{(a)}_0$ and $\mathcal{L}^{(r)}_0$ originate from terms that are at most quadratic in the fields, i.e., those involving $H_0$ and $L_{0,\sigma}$. They contain only $(a)$ or $(r)$ fields, respectively. Specifically, they are given by
\begin{equation}
    \mathcal{L}^{(a)}_0 \rho_2 =  \mathcal{R}[-i H_\textrm{eff,0}^{(a)}] \rho_2  +  2  \sum_\mu \mathcal{D}[L^{(a)}_{0,\mu}]\rho_2 ,\quad
    \mathcal{L}^{(r)}_0 \rho_2 = \mathcal{R}[-i H_\textrm{eff,0}^{(r)}] \rho_2.
\end{equation}
With $H_0^{(\zeta)}$ obtained from Eq.~\eqref{eq:bosonized_hamiltonian}, the effective Hamiltonians for absolute and relative sectors ($\zeta \in \{a,r\}$) have opposite signs for the measurement-induced terms, 
\begin{equation}\label{eq:eff_hamiltonian_zeta}
    H_\textrm{eff,0}^{(\zeta)} = H_0^{(\zeta)} \mp_\zeta \sum_\mu \Big\{M_{0,\mu}^{(\zeta)}K_{0,\mu}^{(\zeta)}  - i \bqty{(M_{0,\mu}^{(\zeta)})^2 - \ev*{(M^{(\zeta)}_{0,\mu})^2}}\Big\},    
\end{equation}
with  $\mp_a = -$ and $\mp_r = +$.  The operators $M_{0,\mu}^{(\zeta)}$ and $K_{0,\mu}^{(\zeta)}$ similarly follow from Eqs.~\eqref{eq:bosonized_jump_operators_lin} by using  relative or absolute  boson fields.
Nonlinear contributions originate from $\delta H$ and $\delta L_\sigma$.  In Eq.~\eqref{eq:2repl}, these terms are collected in the Lindbladian term ($i,j=1,2$)
\begin{equation}\label{eq:deltaL}
    \delta \mathcal{L} \, \rho_2 = -i \sum_i \bqty*{  \delta H^{(i)} , \rho_2 } +  \sum_{ij,\mu} \pqty{ \delta M_\mu^{(i)} \rho_2 \, \delta M_\mu^{(j)} - \frac{(-1)^{i+j}}{2} \Bqty*{  \delta M_\mu^{(i)} \delta M_\mu^{(j)},\rho_2} -  \ev*{\delta M^{(1)}_\mu \delta M^{(2)}_\mu} \rho_2},
\end{equation}
where $\delta H = \delta H_\textrm{Ising} + H_\textrm{flip}$ combines backscattering effects due to the Ising interaction and spin-flip terms, and $\delta M^{(i)}_\mu$ is given by Eq.~\eqref{eq:bosonized_jump_operators_nonlin}.
For clarity, let us now focus on the Zeeman coupling, but using   the Rashba term---or both terms together---instead of the Zeeman coupling   yields identical conclusions at long times 
after averaging over the absolute replica sector. In bosonized language, with the Pauli matrix $\beta_2$ in 
Eq.~\eqref{pauli}, we find the Hamiltonian contributions
\begin{equation}
    \delta H_\textrm{Ising} \sim   - \beta_2 \int dx\, \cos\pqty{\sqrt{8\pi}\varphi_s}, \quad
    H_\textrm{flip} \sim  - \beta_2 \int dx\, \sin\pqty{\sqrt{2\pi}\theta_s} \cos\pqty{\sqrt{2\pi}\varphi_s}. 
\end{equation}
At long times, both terms do not contribute to $\rho_2$ after averaging over the absolute mode.

\section{Average over the absolute mode}

Taking into account that the absolute replica sector heats up for $t\to \infty$ and that $C_{AB}(x)$ involves only fields in the relative sector, we next perform a trace over the absolute modes on both sides of Eq.~\eqref{eq:2repl}.  We thereby arrive at an effective equation of motion for the density operator
$\rho_2^{(r)}={\rm tr}^{(a)}(\rho_2)$ describing the relative sector. 
At long times, the factorization $\rho_2 \simeq \rho^{(a)}_2 \otimes \rho^{(r)}_2$ applies, 
where $\rho^{(a)}_2$ describes the diverging bosonic fluctuations in the absolute sector.  Let us begin by considering contributions due to $\delta {\cal L}$ in Eq.~\eqref{eq:deltaL}. 
We then note that $\tr^{(a)}(\delta H^{(i)} \rho^{(a)}_2) \to 0$ while one has  
\begin{equation} 
     \tr^{(a)}\bqty{\Bqty {  \pqty{ \delta M_\sigma^{(1)} - \delta M_\sigma^{(2)}}^2,\rho^{(a)}_2 \otimes \rho^{(r)}_2} } 
    = \frac{\gamma}{\pi^2}  \Bqty{ 1 -  \cos\pqty{\sqrt{8\pi}\varphi_\sigma^{(r)}}    , \rho^{(r)}_2},
\end{equation}
where $\varphi_\sigma = (\varphi_c + \sigma\varphi_s)/\sqrt{2}$. Using the above product ansatz for $\rho_2$ in a naive manner, the term $\tr^{(a)}[\sum_{ij} \delta M_\mu^{(i)} \rho_2 \delta M_\mu^{(j)}]$ in Eq.~\eqref{eq:deltaL} appears to give a finite contribution. However, a more careful analysis based on the corresponding Keldysh action \cite{buchhold2021effective,mueller2025monitored} shows that this term vanishes upon tracing over the absolute modes. In fact, it depends on inter-contour correlation functions of the type $\exp\Bqty*{ - 2\pi\ev*{(\varphi^{(a)}_+ \pm \varphi^{(a)}_-)^2}}$, where $\pm$ labels the Keldysh forward and backward contours. While usually correlations of the quantum component $\ev*{(\varphi^{(a)}_+ - \varphi^{(a)}_-)^2}$ are guaranteed to vanish by causality \cite{Nazarov2009}, this does not hold true for the replicated monitored theory, where $\ev*{(\varphi^{(a)}_+ \pm \varphi^{(a)}_-)^2}$ grows unboundedly for both signs. We therefore drop the corresponding exponentiated terms as well as the associated contributions to the normalization term. Altogether, after summing over $\sigma$, we find 
\begin{equation}
    \tr^{(a)}( \delta \mathcal{L} \,\rho_2) \simeq  \frac{\gamma}{\pi^2} \int dx\,  
       \Big[\Big\{ \cos\pqty{\sqrt{4\pi}\varphi_c^{(r)}}\cos\pqty{\sqrt{4\pi}\varphi_s^{(r)}} , \rho^{(r)}_2 \Big\}  
     -2 \ev{\cos\pqty{\sqrt{4\pi}\varphi_c^{(r)}}\cos\pqty{\sqrt{4\pi}\varphi_s^{(r)}}} \rho^{(r)}_2 \Big] . 
\end{equation}
Dropping normalization terms for ease of presentation, and collecting the different contributions to 
\begin{equation}
\dot{\rho}^{(r)}_2 = \tr^{(a)}(\dot{\rho}_2) = \mathcal{L}_0^{(r)}\rho^{(r)}_2 + \tr^{(a)}( \delta \mathcal{L} \, \rho_2),
\end{equation}
we obtain Eqs.~\eqref{eq:eom_relative_sector} and \eqref{eq:eff_hamiltonian}. 

\section{Derivation of the RG flow equations} 

We evaluate the long-distance behavior of  correlation functions in the presence of the nonlinear sine-Gordon terms in the nonhermitian effective Hamiltonian. To this end, we introduce a field theory which enables a perturbative one-loop RG calculation.  Mostly dropping the $(r)$ label indicating the relative replica sector, in order to define the field theory, consider the solution to Eq.~\eqref{eq:eom_relative_sector}, 
$\rho^{(r)}(t) = \frac{\ketbra*{\psi(t)}}{ \braket{\psi(t)}},$ 
where we explicitly normalize and use $\ket*{\psi(t)} = e^{-i H_\textrm{eff} t} \ket*{\psi(0)}$
with the nonhermitian sine-Gordon Hamiltonian $H_{\rm eff}$  in Eq.~\eqref{eq:eff_hamiltonian}.
The initial state $\ket*{\psi (0)}$ is not important since the evolution drives any initial state to the same dark state for $t\to \infty$. Using standard techniques, we write the time evolution operator (at long times) as functional integral, 
$e^{-i H_\textrm{eff} t} \to \int \mathcal{D}\varphi\, e^{i S[\varphi]},$
with the real-time action 
$S[\varphi] = S_0[\varphi]+ S_\kappa[\varphi]+ S_\lambda [\varphi].$
Here, we have separated the action into Gaussian ($S_0+S_\kappa$) and nonlinear parts ($S_\lambda$). Rescaling $t \to t/v_F$, with the couplings  in Eq.~\eqref{eq:couplings}, we find  
\begin{eqnarray} \nonumber 
    S_0 &=&\frac{1}{2} \int_{t,x}\, 
    \sum_{\nu=c,s} \bqty{ l_\nu  (\partial_t\varphi_\nu)^2 -  k_\nu   (\partial_x\varphi_\nu)^2 },\quad  
    S_\kappa =  - \kappa   \int_{t,x}\, \bqty{ \partial_x\varphi_c \partial_t\varphi_s + (c\leftrightarrow s)} ,\\
    S_\lambda &=& - i \lambda    \int_{t,x} \cos(\sqrt{4\pi}\varphi_c) \cos(\sqrt{4\pi}\varphi_s) .
\end{eqnarray}
We note that the $\kappa^2$ term in ${\cal H}_{\rm eff,0}$, see Eq.~\eqref{eq:eff_hamiltonian}, cancels out when switching to the 
corresponding action $S_0$. To capture renormalization effects of the Fermi velocity $v_F$, 
we also introduced the couplings $l_\nu$ in $S_0$, with $l_\nu=1$ in the ultraviolet.
We proceed by performing a standard one-loop  momentum-shell renormalization \cite{Gogolin1998}, employing a space-time symmetric frequency and momentum cutoff $\Lambda$. With the flow parameter $\ell$ such that $d\ell=-d\ln\Lambda$, we find the RG equations 
\begin{equation}\label{eq:rg_eqs_full}    
\frac{d \lambda }{d\ell} =  \pqty{2 -\sum_\nu \frac{1}{\sqrt{l_\nu  k_\nu}}} \lambda, \quad
    \frac{d l_\nu }{d\ell}
    = f_t  \lambda^2 ,\quad 
    \frac{d k_\nu}{d\ell}  = - f_x \lambda^2.
\end{equation}
We here assumed $\Im(l_\nu  k_\nu)<0$ to hold during the entire RG flow, which has been confirmed numerically throughout the perturbative RG regime with $|\lambda(\ell)| \lesssim 1$. 
With $l_\nu(0) = 1$, we find $l_c(\ell) = l_s(\ell) = l(\ell)$ for all $\ell$. 
In Eq.~\eqref{eq:rg_eqs_full}, we also introduced the real numbers $f_t$ and $f_x$ through the relation
\begin{equation}\label{eq:f_functions}
    \begin{pmatrix}
        f_t \\ f_x
    \end{pmatrix}
    =    - \frac{1}{2 \Lambda^4  } \int_0^{1/\Lambda} dR\, R^3 
    \int_0^{2\pi} d\chi \begin{pmatrix}
        \cos^2 \chi \\ \sin^2\chi
    \end{pmatrix}   \int_0^{2\pi} d\vartheta \sum_{\nu=c,s}   
    \frac{e^{  -i  R \cos(\chi + \vartheta)}}{l_\nu \cos^2\vartheta -  k_\nu  \sin^2\vartheta}. 
\end{equation}
For $\Lambda \sim 1$, we find $f_t \simeq f_x$ and thus set $f_t = f_x = f$ in Eq.~\eqref{eq:RG_eqs}, with $|f|\sim {\cal O}(1)$.  
Further, in the perturbative regime, $l(\ell)$ does not significantly depart from its initial value $l(0) = 1$, and we therefore drop this factor in Eq.~\eqref{eq:RG_eqs}.
For the results shown in Fig.~2(a) of the main text, we have numerically integrated the full RG equations \eqref{eq:rg_eqs_full}, with $f_{t,x}$ in Eq.~\eqref{eq:f_functions} and setting $\Lambda = 1$. To determine the phase boundary, we integrate Eq.~\eqref{eq:rg_eqs_full} from $\ell = 0$ to $\ell_\textrm{max} = 50$, using as initial data $\lambda(0) = \lambda$, $l_\nu (0) = 1$, and $k_\nu(0)=k_\nu$, see Eq.~\eqref{eq:couplings}. The strong vs weak coupling phases are identified through the criterion $\abs{\lambda(\ell_\textrm{max})/\lambda(0)} \gtrless 1$.  

\section{Current operator}

In addition to the usual Hamiltonian contribution, the current operator can also receive contributions originating from  quantum jumps described by $L_{j,
\sigma}$ in Eq.~\eqref{eq:lattice_jump_operators}, see Ref.~\cite{antonic2025motion}. To identify the corresponding bosonized current operator, we consider the (measurement-averaged) Heisenberg picture evolution of $Q_\sigma(x) = \int_{-\infty}^x dx'\, \varrho_\sigma(x')$, where $\varrho_\sigma(x) = \frac{1}{\sqrt{\pi}} \partial_x \varphi_\sigma(x) + (2k_F\textrm{-terms})$
is the local density variation of spin-$\sigma$ fermions. By charge conservation, we identify the local current as $j_\sigma(x) = - \dot{Q}_\sigma(x)$.  (In this expression, we have assumed that spin-flip processes are irrelevant, as is the case after averaging over the absolute replica sector. In the presence of spin-flip processes, only the charge current $\sum_\sigma j_\sigma(x)$ is well-defined.) The full spin-resolved current density, including measurement-induced current operators, may then be evaluated by using the Lindblad equation, resulting in
\begin{equation}\label{eq:dotQ}
\dot{Q}^{}_{\sigma}(x) = \mathcal{L}^\dagger Q^{}_{\sigma}(x),\quad 
\mathcal{L}^\dagger \mathcal{O} = i[H,\mathcal{O}] + \sum_\mu \left(L^\dagger_\mu \mathcal{O} L_\mu^{} - \tfrac{1}{2} \{L^\dagger_\mu  L_\mu^{} , \mathcal{O}\}\right).
\end{equation}
Dropping rapidly oscillating contributions, i.e., setting $\varrho_\sigma(x) \equiv  \frac{1}{\sqrt{\pi}} \dot{\varphi}_\sigma(x)$ \cite{Gogolin1998}, we arrive at
\begin{equation}\label{current1}
  j_\sigma(x) = - \frac{v_F}{\sqrt{\pi}} \partial_x\theta_\sigma(x)  + \frac{v_F \kappa}{\sqrt{\pi}} \sigma \partial_x\varphi_\sigma(x) + \sigma \frac{j_0}{2},
\end{equation}
where $j_0 = v_F \kappa \varrho_0$ depends on the average fermion density $\varrho_0$ in the system. 
The first term in Eq.~\eqref{current1} comes from the Hamiltonian contribution, while the two other terms are due to measurement-induced currents. In fact, with $\kappa$ in Eq.~\eqref{eq:couplings}, the spin-dependent average current $\sigma j_0/2$ in Eq.~\eqref{current1} depends linearly on the monitoring strength $\gamma$ 
and highlights the quantum active motion of particles exposed to this monitoring protocol.
Forming charge and spin current densities via $j_c = \sum_\sigma j_\sigma$ and $j_s = \sum_\sigma \sigma j_\sigma$, we obtain Eq.~\eqref{eq:current_ops}.
Notably, current operators of  similar form can be obtained by gauging the $U(1)$ charge and spin symmetries of the effective Hamiltonians  in Eq.~\eqref{eq:eff_hamiltonian_zeta}. With our conventions and $\zeta \in \{a,r\}$, minimal coupling is implemented through the replacement 
$\partial_x \theta_\nu^{(\zeta)}(x)  \to  \partial_x \theta_\nu^{(\zeta)}(x) +  \sqrt{\frac{2}{\pi}} \mathcal{A}_\nu(x)$.
This gives the minimal coupling currents 
\begin{align} 
    j_{\textrm{mc};\nu}^{(\zeta)}(x) =  - \frac{\delta H_\textrm{eff,0}^{(\zeta)} }{\delta {\cal A}_\nu(x)} 
     = - \frac{v_F}{\sqrt{\pi/2}} \bqty{ \partial_x\theta^{(\zeta)}_\nu   \mp_\zeta  \kappa \partial_x\varphi^{(\zeta)}_{\overline{\nu}}}, 
\end{align}
where again $\mp_a = -$ and $\mp_r = +$. Note the difference in relative sign to Eq.~\eqref{eq:current_ops} for $\zeta = r$. This has important implications for the RG flow of connected correlation functions involving   $j^{(r)}_\nu$: Only the relative replica sector $U(1)$ charge densities and the associated minimal coupling currents remain normalized under the RG flow, i.e.,  $\varrho_\nu^{(r)} (x) \big\vert_\textrm{UV} =\varrho_\nu^{(r)} \big\vert_\textrm{IR} + \ldots$ and
$j_{\textrm{mc};\nu}^{(r)}(x) \big\vert_\textrm{UV} =  j_{\textrm{mc};\nu}^{(r)}(x) \big\vert_\textrm{IR} + \ldots$,
where  ellipses refer to irrelevant operators. However, we are interested in the connected correlations of the current operator defined in Eq.~\eqref{eq:current_ops}. Using Eqs.~\eqref{eq:correlation_1} and \eqref{eq:correlation_2}, this translates to correlations involving  
    $j_{\nu}^{(r)}= j_{\textrm{mc};\nu}^{(r)} + 2 v_F \kappa \varrho_{\overline{\nu}}^{(r)}$.
Under the RG flow, this maps to 
$j_{\nu}^{(r)} \vert_\textrm{UV} = j_{\textrm{mc};\nu}^{(r)}\vert_\textrm{IR}  + 
2 v_F \kappa \varrho_{\overline{\nu}}^{(r)} \vert_\textrm{IR} + \ldots,$ and by
denoting the parameters of the renormalized action as $v_F'$ and $\kappa'$, we thus find 
\begin{equation}
    j_{\nu}^{(r)}   \big\vert_\textrm{UV} = -  \sqrt{\frac{2}{\pi}} \left[ v_F'  \partial_x\theta_c   +   (2v_F \kappa - v_F'\kappa')  \partial_x\varphi_s \right]_\textrm{IR} + \ldots 
\end{equation} 
Since neither $v_F$, as captured by the RG flow of $l(\ell)$, nor $\kappa$ are significantly renormalized during the perturbative RG flow,
we set $2v_F \kappa - v_F'\kappa' \simeq v_F \kappa$. In the main text, we thereby neglect the  weak renormalization of measurement-induced currents.

\section{Correlations at the Gaussian fixed points}

We compute the steady-state correlation functions $C_{AB}(x)$ at the fixed points describing either the weak-coupling limit (with $\lambda\to 0$) 
or the strong-coupling limit (with $|\lambda|\to \infty$).  Both fixed points are described by a Gaussian theory $H_{\rm gfp}$ in terms of boson mode operators, see also Ref.~\cite{buchhold2021effective}.
This fact allows us to obtain exact results for $C_{AB}(x)$.
Suppressing the $(r)$ superscript again, we expand the bosonic fields $\varphi_\nu(x)$ and $\theta_\nu(x)$ by using bosonic mode operators $b_{\nu,q}$  with the algebra $[b^{}_{\nu,q},b^\dagger_{\nu',q'}] = \delta_{\nu\nu'}\delta_{qq'}$ \cite{Gogolin1998}. Specifically, we have 
\begin{equation}
    \varphi_{\nu}(x) =  \sum_q \sqrt{\frac{v_F}{2L\abs{\omega_{\nu,q}}}}  e^{iqx} (b^\dagger_{\nu,q} + b^{}_{\nu,-q}),\quad
    \partial_x\theta_{\nu}(x) = i\sum_q \sqrt{\frac{\abs{\omega_{\nu,q}}}{2 L v_F}} e^{iqx} (b^\dagger_{\nu,q} - b^{}_{\nu,-q}),
\end{equation}
with the complex dispersion relation
\begin{equation}\label{disp}
    \omega_{\nu,q}^2 =  \tilde k_\nu v_F^2 q^2 - i m^2_\lambda, \quad \tilde k_\nu\equiv k_\nu+\kappa^2.
\end{equation}
Here, we have $m_\lambda=0$ at the weak-coupling fixed point, while the nonhermitian sine-Gordon term yields a finite value for the ``mass'' $m_\lambda$ at strong coupling. 
We note that for  $\gamma>0$, we have $\Im(\omega_{\nu,q}) < 0$, see Eq.~\eqref{eq:couplings}. 
With bosonic spinor operators $B_q$, the fixed-point Hamiltonian has the Gaussian form 
\begin{equation}
    H_{\textrm{gfp}} = \frac{1}{2} \sum_q B_q^\dagger h_q^{} B^{}_q,
    \quad B_q = \left(b^{}_{c,q},b^{}_{s,q},b^\dagger_{c,-q},b^\dagger_{s,-q}\right)^\top,\quad
\label{eq:sp_hamiltonian}
    h_q =\begin{pmatrix}
         h_{c,q} & 0 \\ 0 & h_{s,q}
    \end{pmatrix}_{\sigma} + v_F q (\kappa_{q,+}  \sigma_1 \tau_3 - \kappa_{q,-} \sigma_2 \tau_2 ),
\end{equation}
where the Hamiltonian matrix $h_q$ and the operators $B_q$ are defined in a combined charge-spin and Nambu (particle-hole) space. In particular, the $h_{\nu,q}$
in Eq.~\eqref{eq:sp_hamiltonian} correspond to the Nambu matrices 
\begin{equation}
    h_{\nu,q} = \frac{\abs{\omega_{\nu,q}}}{2} \bqty{\pqty{e^{i\eta_{\nu,q}} +1 } \tau_0+\pqty{e^{i\eta_{\nu,q}}-1 } \tau_1},\quad
\eta_{\nu,q} = 2 \,\textrm{arg}\, (\omega_{\nu,q}).
\end{equation} 
Here and below, the subscripts $\sigma$ and $\tau$ refer to charge-spin ($\nu,\nu'=c,s)$ and Nambu space, respectively. Similarly, Pauli matrices $\sigma_i$ ($\tau_i)$ act in charge-spin (Nambu) space.  In Eq.~\eqref{eq:sp_hamiltonian}, we also used the couplings 
\begin{equation}
    \kappa_{q,\pm} = \frac{\kappa}{2} \left(\sqrt{\left|\frac{\omega_{c,q}}{\omega_{s,q}}\right|} \pm \sqrt{\left|\frac{\omega_{s,q}}{\omega_{c,q}}\right|}\right).
\end{equation}
To identify the dark state $\ket{0}$, we perform a Bogoliubov transformation, $B_q = V_q \tilde{B}_q$, in order to bring $\tilde{h}_q = V_q^\dagger h_q V_q$ into lower triangular form in Nambu space. Note that $V_q$ is not unitary but rather satisfies $V^\dagger_q\tau_3 V_q = \tau_3$. With indices referring to the Nambu basis, the Hamiltonian is then given by
\begin{equation}
    H_{\textrm{gfp}} = \sum_q \pqty{\tilde{b}^\dagger_q \tilde{h}_{q,11} \tilde{b}^{}_q + \frac{1}{2} \tilde{b}^{}_{-q} \tilde{h}_{q,21}\tilde{b}^{}_q}. 
\end{equation}
The dark state is identified as the unique state annihilated by the $\tilde{b}^{}_q$ bosons,   $\tilde{b}^{}_q\ket{0} = 0$. 

Steady-state equal-time correlation functions $C_{AB}(x)$ are simply correlations in the state $\ket{0}$. 
We are interested in correlation functions of charge and spin currents, $j_\nu$, and of the respective densities, $\varrho_{\nu'}$.  Using Eq.~\eqref{eq:current_ops} for $j_\nu$ (and similarly for $\varrho_{\nu}$, see main text),  these correlation functions can be decomposed as
\begin{eqnarray} \label{eq:density_and_current_correlations}
    C_{\varrho^{}_\nu\varrho_{\nu'}}(x) &=& \frac{2}{\pi} F^{++}_{\nu\nu'}(x),\quad
    C_{\varrho^{}_\nu j_{\nu'}}(x) = -\frac{2}{\pi}v_F \bqty{ F^{+-}_{\nu\nu'}(x) - \kappa F^{++}_{\nu\overline{\nu}'}(x) },\\ \nonumber
    C_{j^{}_\nu \varrho_{\nu'}}(x) &=& -\frac{2}{\pi}v_F \bqty{ F^{-+}_{\nu\nu'}(x) - \kappa F^{++}_{\overline{\nu}\nu'}(x) },\quad 
    C_{j^{}_\nu j_{\nu'}}(x) = \frac{2}{\pi} v_F^2 \big[F^{--}_{\nu\nu'}(x) + \kappa^2 F^{++}_{\overline{\nu}\overline{\nu}'}(x) 
    - \kappa \pqty{F^{-+}_{\nu\overline{\nu}'}(x) +F^{+-}_{\overline{\nu}\nu'}(x)} \big],
\end{eqnarray}
with the correlation functions ($\alpha,\alpha'=\pm$)
\begin{align}\label{eq:F_correlation_fn_0}
    F^{\alpha\alpha'}_{\nu\nu'}(x) =&\ \bra{0}\partial_x \chi_{\alpha,\nu}(x)\partial_x \chi_{\alpha',\nu'}(0) \ket{0}.
\end{align}
We recall that $\chi_{+,\nu} = \varphi_{\nu}$ and $\chi_{-,\nu} = \theta_\nu$. 
In Eqs.~\eqref{eq:density_and_current_correlations}, terms proportional to $\kappa$ or $\kappa^2$ originate from  measurement-induced currents. We note that the constant measurement-induced spin current $j_0=v_F\kappa\varrho_0$ does not affect connected correlation functions. 
Taking the $L\to \infty$ limit,  Eq.~\eqref{eq:F_correlation_fn_0} can be written as
\begin{equation}\label{eq:F_correlation_fn_1}
    F^{\alpha\alpha'}_{\nu\nu'}(x) = \int \frac{dq}{8\pi} e^{iqx}  q \pqty{\frac{v_Fq}{\abs{\omega_{\nu,q}}}}^{\frac{\alpha}{2}}
    \pqty{\frac{v_Fq}{\abs{\omega_{\nu',q}}}}^{\frac{\alpha'}{2}}   \bqty{\tr\pqty{V_q^\dagger M_{\nu\nu'}^{\alpha\alpha'} V_q} -2 \delta_{\nu\nu'}(1-\delta_{\alpha\alpha'})}, \quad
    M_{\nu\nu'}^{\alpha\alpha'} = \ketbra{\nu}{\nu'}_\sigma \otimes \begin{pmatrix}
        1 & \alpha' \\
        \alpha & \alpha\alpha'
    \end{pmatrix}_\tau.
\end{equation}

We can argue on general grounds that there are no  correlations between the charge fields $\chi_{c}$ and the spin fields $\chi_s$ if $g_c = g_s$. In this case, $h_{c,q} = h_{s,q} \equiv h_{0,q}$ and $\kappa_{q,-} = 0$, such that one can write the single-particle Hamiltonian as $h_q =  \sigma_0 h_{0,q} + \kappa q \sigma_1 \tau_3$. Note that $\sigma_1 = \pm 1$ corresponds to spins $\uparrow$ and $\downarrow$, respectively, i.e., the system decouples into two independent non-hermitian Luttinger liquids. In particular, due to the absence of spin-structure in the first term, one can choose $V_q \propto \sigma_0$. Further, due to $V^\dagger_q \tau_3 V_q = \tau_3$, the $\kappa$-term does not affect the transformation $V_q$. As a consequence, the correlation functions $F$ are independent of $\kappa$ and vanish for cross-correlations between the spin-$\sigma$ sectors, and also for cross-correlations between the spin and charge sector. However, correlations between the spin/charge density and the charge/spin current persist even for $g_c = g_s$ due to the $\kappa$-term in Eq.~\eqref{eq:current_ops}.

\subsection{Weak-coupling fixed point}

We now proceed by evaluating the correlation functions $F$. Consider first the massless case, $m_\lambda = 0$, corresponding to the fixed point of the quasi-long-range quantum active phase. Here, we have
$\abs{\omega_{\nu,q}} = \abs{  \sqrt{ k_\nu }v_F q},$
and the only nontrivial momentum dependence of $h_q$ stems from the sign of $q$ in the $\kappa$ term in Eq.~\eqref{eq:sp_hamiltonian}. Thus, the transformation $V_q$ depends only on the sign of $q$, allowing us to decompose
\begin{equation}\label{eq:sigma_delta}
    \tr\pqty{V_q^\dagger M_{\nu\nu'}^{\alpha\alpha'} V_q - M_{\nu\nu'}^{\alpha\alpha'}\tau_3 } = \tilde\Sigma_{\nu\nu'}^{\alpha\alpha'} + \textrm{sgn}(q) \tilde\Delta_{\nu\nu'}^{\alpha\alpha'},
\end{equation}
in terms of the momentum-independent coefficients
\begin{equation}      
\tilde\Sigma_{\nu\nu'}^{\alpha\alpha'} = \frac{1}{2} \sum_{\pm} \tr\pqty{V_\pm^\dagger M_{\nu\nu'}^{\alpha\alpha'} V_\pm }- \delta_{\nu\nu'}(1-\delta_{\alpha\alpha'}),\quad
    \tilde\Delta_{\nu\nu'}^{\alpha\alpha'} = \frac{1}{2} \sum_{\pm} (\pm)\tr\pqty{V_\pm^\dagger M_{\nu\nu'}^{\alpha\alpha'} V_\pm}.
\end{equation}
Here, we defined $V_+ = V_{q>0}$ and $V_- = V_{q<0}$. They are not independent: The relation $V_- = \tau_1 V_{+}^* \tau_1$ implies
\begin{equation}\label{eq:link_positive_and_negative}
    \tr\pqty{  V^\dagger_- M^{\alpha\alpha'}_{\nu\nu'} V_- } = \alpha\alpha' \tr\pqty{ V^\dagger_+  M^{\alpha'\alpha}_{\nu'\nu}    V_+}.
\end{equation}
Equation \eqref{eq:F_correlation_fn_1} then becomes
\begin{equation} \label{eq:F_correlation_fn_2}
    F^{\alpha\alpha'}_{\nu\nu'}(x) = \abs*{\tilde k_\nu}^{-\frac{\alpha}{4}}\abs*{\tilde k_{\nu'}}^{-\frac{\alpha'}{4}}\int \frac{dq}{8\pi} e^{iqx}  q    \Bqty{[\textrm{sgn}(q)]^{\delta_{\alpha\alpha'}} \tilde\Sigma_{\nu\nu'}^{\alpha\alpha'} + [\textrm{sgn}(q)]^{1 - \delta_{\alpha\alpha'}} \tilde\Delta_{\nu\nu'}^{\alpha\alpha'} } .
\end{equation}
We observe that the spatial dependence of this correlation function stems either from the Fourier transform of $\abs{q}$, giving rise to quasi-long-range algebraic decay $\propto 1/x^2$, or from $q$, which instead gives ultra-local $\partial_x \delta(x)$ behavior. We are interested only in the asymptotic tails. Thus, for $\alpha = \alpha'$, corresponding to $\varphi$--$\varphi$ or $\theta$--$\theta$ correlations, we keep only the $\tilde\Sigma$-term, while for $\alpha'=\overline{\alpha}\equiv - \alpha$, corresponding to $\varphi$--$\theta$ correlations, we retain only the $\tilde\Delta$-term. Using Eq.~\eqref{eq:link_positive_and_negative} the corresponding coefficients can be expressed as 
\begin{equation}   
    \tilde\Sigma_{\nu\nu'}^{\alpha\alpha} = \frac{1}{2} \tr\bqty{V^\dagger_+ X_{\nu\nu'} \pqty{\tau_0 + \alpha \tau_1}  V_+},\quad
    \tilde\Delta_{\nu\nu'}^{\alpha\overline{\alpha}} = \frac{1}{2} \tr\bqty{V^\dagger_+ \pqty{X_{\nu\nu'}  \tau_3 + \alpha  Y_{\nu\nu'} \tau_2} V_+}
    = \tilde\Delta_{\nu'\nu}^{\overline{\alpha}\alpha},
\end{equation}
where we introduced the shorthand notations
$X_{\nu\nu'} = \ketbra{\nu}{\nu'} + \textrm{h.c.}$ and $Y_{\nu\nu'} = -i (\ketbra{\nu}{\nu'} - \textrm{h.c.})$.
We find 
\begin{equation}\label{eq:vanishing}
\tilde\Sigma_{\nu\overline{\nu}}^{\alpha\alpha} = 0,\quad\tilde\Delta_{\nu\nu}^{\alpha\overline{\alpha}} = 0.
\end{equation}
The first relation in Eq.~\eqref{eq:vanishing} implies that only $\varphi_\nu$--$\varphi_\nu$ and 
$\theta_\nu$--$\theta_\nu$ correlations within the same charge-spin sector exhibit algebraic tails. Similarly, the second relation means that only different charge-spin sectors produce quasi-long-range behavior in $\varphi_\nu$--$\theta_{\overline{\nu}}$ cross-correlations. 
Before proceeding, let us outline the proof of Eq.~\eqref{eq:vanishing}.

First, we simplify the relevant matrix elements by noting that $X_{\nu \overline{\nu}} = \sigma_1$, $X_{\nu\nu} = \sigma_0 \pm_\nu \sigma_3$ with $\pm_c = +$ and $\pm_s = -$, and $Y_{\nu\nu} = 0$. This gives
\begin{equation}  
  \tilde\Sigma_{\nu\overline{\nu}}^{\alpha\alpha} = \frac{1}{2} \tr\bqty{V^\dagger_+ \sigma_1 \pqty{\tau_0 + \alpha \tau_1}  V_+},\quad
    \tilde\Delta_{\nu\nu}^{\alpha\overline{\alpha}} =\frac{1}{2} \tr\bqty{V^\dagger_+ (\sigma_0 \pm_\nu \sigma_3)  \tau_3  V_+}. 
\end{equation}
We show below that the transformation $V_+$ can be chosen as $V_+ = \exp{v_1 \tau_1 + v_2 \tau_2}$ with $v_i = v_{i0} \sigma_0 + v_{i2} \sigma_2 + v_{i3} \sigma_3,\ v_{ij}\in \mathbb{R}$. Importantly, $\sigma_1$ does not feature in the exponent. By expanding the exponential order by order, it is straightforward to check that 
$V_+ V^\dagger_+ \in \textrm{span}(\sigma_{i \in \{0,2,3\}}\tau_{j \in \{1,2,3\}},\sigma_1\tau_3),$
i.e., $\sigma_1$ and $\tau_3$ feature only as product in $V_+ V^\dagger_+$. Using cyclicity of the trace and orthogonality, $\tr(\sigma_{i} \tau_{j} \sigma_{i'}\tau_{j'}) = 4\delta_{ii'}\delta_{jj'}$, this immediately implies Eq.~\eqref{eq:vanishing}. 
To see why one may write $V_+$ in this form, first note that $V_+^\dagger \tau_3 V_+ = \tau_3$ implies the general form 
$V_+ = \exp{i(v_0 \tau_0 + v_3 \tau_3) + v_1 \tau_1 + v_2 \tau_2},$
where $v_i$ are (thus far unconstrained) hermitian $2\times 2$ matrices. However, this parametrization is redundant: the condition
$[V_+^\dagger h_{q>0} V_+]_{12} = 0$
corresponds to 8 real equations, while such $V_+$ has 16 real parameters. This redundancy corresponds to the freedom to perform independent unitary spin rotations in the $\tau_3 = \pm 1$ sectors, which leave $[V_+^\dagger h_{q>0} V_+]_{12} = 0$ invariant. Such a spin rotation amounts to a redefinition $V \to VU$ with $U = \exp{i(u_0 \tau_0 + u_3 \tau_3)}$, $u_i$ hermitian. One can use this freedom to eliminate the $\tau_0$ and $\tau_3$ terms in the exponent of $V_+$: choose $U = \exp{-i(v_0 \tau_0 + v_3 \tau_3) + \ldots}$, where $\ldots$ corresponds to terms of order $v^3$ and higher. Applying the Baker-Campbell-Hausdorff formula to combine the exponentials of $V$ and $U$, one finds that with this choice of $U$, only $\tau_{1,2}$-terms appear at order $v^2$, while $\tau_{0,3}$-terms feature at order $v^3$. To eliminate the latter, we include an order-$v^3$ counter-term in the exponent of $U$ (entering the $\ldots$). This term gives rise to new $\tau_{0,3}$-terms only at order $v^5$, which, together with $\tau_{0,3}$-terms at order $v^5$ originating from the leading term in the exponent of $U$, can be accounted for by a $v^5$-contribution to the exponent of $U$. 
In this way, one can eliminate $\tau_{0,3}$-terms in $V_+$. This shows that one can choose 
\begin{equation}\label{eq:V_choice}
    V_+ = \exp{v_1 \tau_1 + v_2 \tau_2}.
\end{equation}
It remains to show that $\tr(v_{1,2}\sigma_1) = 0$. To this end, consider $V_+^\dagger h_{q>0} V_+$ to leading order in $v$ and demand that the Nambu-$12$ block vanishes. This gives 
$\pqty{h + \Bqty{v_1 \tau_1 + v_2 \tau_2 , h}}_{12}=0.$
Multiplying this equation by $\sigma_1$ and taking the trace gives 
$\sum_\nu \frac{\abs{\omega_{\nu,q}}}{2}  \pqty{e^{i\eta_{\nu,q}} +1 } (v_{11} - i v_{21}) = 0,$
which can only be satisfied if $v_{11} = v_{21} = 0$. It is straightforward to check that this pattern persists to higher orders. 

In the end, Eq.~\eqref{eq:F_correlation_fn_2}  simplifies to
\begin{equation} \label{eq:F_correlation_fn_final}
    F^{\alpha\alpha}_{\nu\nu'}(x) = - \frac{1}{4\pi x^2}  \abs{ \tilde k_\nu \tilde k_{\nu'} }^{-\frac{\alpha}{4}} \tilde\Sigma_{\nu\nu'}^{\alpha\alpha} + \ldots,\quad
    F^{\alpha\overline{\alpha}}_{\nu\nu'}(x) = - \frac{1}{4\pi x^2}  \abs{\frac{ \tilde k_\nu }{ \tilde k_{\nu'}} }^{-\frac{\alpha}{4}} \tilde\Delta_{\nu\nu'}^{\alpha\overline{\alpha}} + \ldots,
\end{equation}
where the ellipses denote short-range terms neglected below.     
Collecting the above results, at the $\lambda\to 0$ fixed point, the equal-time steady-state correlation functions $C_{AB}(x)$ in Eq.~\eqref{eq:density_and_current_correlations}  are given by Eq.~\eqref{eq:correlationsfinal},
involving only the six real-valued coefficients ($\nu=c,s$)
\begin{equation}
    \Sigma^+_\nu =  \frac{1}{2\pi^2}\abs*{{\tilde k}_\nu}^{ -\frac{1}{2}} \tilde\Sigma_{\nu\nu}^{++}, \quad
    \Delta_{\nu\overline{\nu}} = -\frac{1}{2\pi^2}\abs*{\tilde{k}_{\overline{\nu}}/\tilde{k}_\nu}^{\frac{1}{4}} 
    \tilde\Delta_{\nu\overline{\nu}}^{+-} 
     + \kappa \Sigma_{\nu}^{+} ,\quad
   \Sigma^-_\nu =  \frac{1}{2\pi^2}\abs*{\tilde k_\nu}^{\frac{1}{2}} \tilde\Sigma_{\nu\nu}^{--}  
   + 2\kappa \Delta_{\overline{\nu}\nu} - \kappa^2 \Sigma_{\overline{\nu}}^+  .
\end{equation} 
To obtain their values as shown in Fig.~2 of the main text, we have numerically determined the transformation $V_+$. 

\subsection{Strong-coupling fixed point}

We next consider the massive case $m_\lambda \neq 0$ in Eq.~\eqref{disp}, where a Gaussian fixed point describes the strong-coupling limit $|\lambda|\to \infty$. As opposed to the massless case, the transformation $V_q$ now also depends on the magnitude of $q$, making it slightly more cumbersome to find the spatial asymptotics of $F_{\nu\nu'}^{\alpha\alpha'}(x)$.
We can make analytical progress by treating the $\kappa$ term perturbatively. To this end, we first find rotations $V_{\nu,q}$ that bring $h_{\nu,q}$ into lower triangular form. It is straightforward to see that this is achieved by the Bogoliubov transformation $V_{\nu,q} = \exp{v_{\nu,q}\tau_2}$ with $v_{\nu,q}$ defined by  
\begin{equation}\label{eq:block_trafo}
    \sinh(2v_{\nu,q}) =  \frac{\Im(\omega_{\nu,q})}{\Re(\omega_{\nu,q})},\quad \tilde{h}_{\nu,q} \equiv V_{\nu,q}^\dagger h_{\nu,q} V_{\nu,q} = \omega_{\nu,q} \tau_0 +  \abs{\omega_{\nu,q}} \pqty{e^{i\eta_{\nu,q}} - 1} \tau_-.
\end{equation}
Applying $V^{(0)}_q = \textrm{diag}(V_{c,q},V_{s,q})$ to $h_q$ brings the first term in Eq.~\eqref{eq:sp_hamiltonian} into lower triangular form, leaving only a contribution due to the $\kappa$-term, 
\begin{equation}
    \bqty{(V^{(0)}_q)^\dagger  h_{q}  V^{(0)}_q}_{12} = v_F q  \sigma_2 \bqty{- \kappa_+  \sinh(v_{c,q}-v_{s,q}) + i\kappa_- \cosh(v_{c,q}+v_{s,q})}.
\end{equation}
We next apply a second Bogoliubov transformation $W_q$ close to unity that eliminates the $\kappa$-term in the Nambu-$12$ block to leading order. Explicitly, we demand
\begin{equation}
    [W_q^\dagger \textrm{diag}(\tilde{h}_{c,q},\tilde{h}_{s,q}) W_q]_{12} = - \bqty{(V^{(0)}_q)^\dagger  h_{q}  V^{(0)}_q}_{12}. 
\end{equation}
This is achieved by
\footnote{
$\textrm{diag}(\tilde{h}_{c,q},\tilde{h}_{s,q})$ contains $\sigma_0$ and $\sigma_3$ only, so we expect $W$ to contain $\sigma_1$ or $\sigma_2$. As argued in the discussion around Eq.~\eqref{eq:V_choice}, there cannot be a $\sigma_1$ contribution.}
$W = 1 + (w_1 \tau_1 + w_2\tau_2) \sigma_2,$
with the coefficients 
\begin{align}\nonumber
    w_1 =&\ \frac{v_F q }{\abs{\sum_\nu \omega_\nu}^2} \sum_\nu \bqty{\kappa_+ \sinh(v_{c,q}-v_{s,q}) \Re(\omega_\nu) -  \kappa_- \cosh(v_{c,q}+v_{s,q}) \Im(\omega_\nu)},\\
    w_2 =&\ \frac{v_F q }{\abs{\sum_\nu \omega_\nu}^2} \sum_\nu \bqty{\kappa_+ \sinh(v_{c,q}-v_{s,q}) \Im(\omega_\nu) +  \kappa_- \cosh(v_{c,q}+v_{s,q}) \Re(\omega_\nu)}. 
\end{align}
Consider now the correlation functions in Eq.~\eqref{eq:F_correlation_fn_1}. To proceed, we use the approximate expression $V_q \simeq V^{(0)}_q W_q$, giving
\begin{equation}\label{eq:traces_perturbation}
    \tr\pqty{V_{q}^\dagger M_{\nu\nu'}^{\alpha\alpha'} V_{q} } \simeq \tr\pqty{(V^{(0)}_{q})^\dagger M_{\nu\nu'}^{\alpha\alpha'}V^{(0)}_{q} } + \tr\pqty{\Bqty{(w_1 \tau_1 + w_2\tau_2) \sigma_2, (V^{(0)}_{q})^\dagger M_{\nu\nu'}^{\alpha\alpha'}V^{(0)}_{q}} }.
\end{equation}
We derive first the correlations originating from the first term. $V^{(0)}_{q}$ does not mix charge and spin sectors, and thus at this order there are no such cross-correlations. Explicitly, 
\begin{align}
    F^{\alpha\alpha'}_{\nu\nu'}(x)\Big\vert^{(0)} \equiv&\ \int \frac{dq}{8\pi} e^{iqx}  q \pqty{\frac{v_Fq}{\abs{\omega_{\nu,q}}}}^{\frac{\alpha}{2}}
    \pqty{\frac{v_Fq}{\abs{\omega_{\nu',q}}}}^{\frac{\alpha'}{2}}    \Bqty{\tr\pqty{(V^{(0)}_{q})^\dagger M_{\nu\nu'}^{\alpha\alpha'}V^{(0)}_{q} } - 2 \delta_{\nu\nu'}(1-\delta_{\alpha\alpha'})} \\
    =&\ \delta_{\nu\nu'} \int \frac{dq}{8\pi} e^{iqx}  q \pqty{\frac{v_Fq}{\abs{\omega_{\nu,q}}}}^{\frac{\alpha+\alpha'}{2}}   \Bqty{\tr\pqty{(V^{(0)}_{q})^\dagger M_{\nu\nu}^{\alpha\alpha'} V^{(0)}_{q} } - 2 (1-\delta_{\alpha\alpha'})}
\end{align}
For $\alpha' = -\alpha$, this gives ultra-short range behavior,
\begin{align}
    F^{\alpha\overline{\alpha}}_{\nu\nu}(x)\Big\vert^{(0)} = 
    \int \frac{dq}{8\pi} e^{iqx}  q  \Bqty{\tr\pqty{V_{\nu,q}^\dagger  \pqty{\tau_3 -i\alpha \tau_2} V_{\nu,q}  } - 2 } \propto \partial_x \delta(x).
\end{align}
For $\alpha' = \alpha$, one instead finds  
\begin{align}
    F^{\alpha\alpha}_{\nu\nu}(x)\Big\vert^{(0)} = \int \frac{dq}{8\pi} e^{iqx}  q \pqty{\frac{v_Fq}{\abs{\omega_{\nu,q}}}}^{\alpha}    \tr\pqty{V_{\nu,q}^\dagger  \pqty{\tau_0 + \alpha \tau_1} V_{\nu,q} } = \int \frac{dq}{4\pi} e^{iqx}  q \pqty{\frac{v_Fq}{\abs{\omega_{\nu,q}}}}^{\alpha}     \frac{\abs{\omega_{\nu,q}}}{\Re(\omega_{\nu,q})}.
\end{align}
The behavior of $F^{\alpha\alpha}_{\nu\nu}(x)$ at large $|x|$ is controlled by the branch points of the integrand which occur at
   $v_F q \in \Bqty{ \pm \sqrt{\frac{i m_\lambda^2}{k_\nu}} , \pm \sqrt{\frac{ - i (m^*_\lambda)^2}{k^*_\nu}}},$
leading to an exponential decay, 
\begin{equation}
    F^{\alpha\alpha}_{\nu\nu}(x)\Big\vert^{(0)} = \frac{f^\alpha_\nu(x)}{(\abs{x}/\xi_\nu)^{3/2}} e^{- \abs{x}/\xi_\nu},\quad \xi_\nu = v_F\abs{\Im\pqty{\sqrt{\frac{i m_\lambda^2}{k_\nu}}}}^{-1},
\end{equation}
where the $\alpha$-dependent prefactor  $f^\alpha_\nu(x)$ oscillates on a scale set by the wavevector $\abs*{\Re\pqty*{\sqrt{i m_\lambda^2 / k_\nu }}}/v_F$. 

Next, we analyze which of the thus far vanishing correlations become finite when the leading correction due to the $\kappa$-term is included. As $\ln W$ is off-diagonal in charge-spin space, see Eq.~\eqref{eq:traces_perturbation}, it is clear that only $\nu'=-\nu$ correlations receive a contribution. The set of branch points now involves the branch points due to $\omega_{c,q}$ and $\omega_{s,q}$. The decay is set by the point closest to the real axis, resulting in $F^{\alpha\alpha'}_{\nu\overline{\nu}} \sim \kappa e^{-\abs{x}/\xi_{cs}}$ with $\xi_{cs} = \textrm{max}\Bqty{\xi_c,\xi_s}$. 
To conclude, in this phase we find short-ranged correlations, where the length scale governing the exponential decay is described by Eq.~\eqref{eq:correlation_length}.

\end{document}